\documentclass[prd, aps, superscriptaddress, showpacs]{revtex4}

\usepackage{latexsym, graphicx, color}

\begin{document}

\title{Quantum teleportation between moving detectors in a quantum field}
\author{Shih-Yuin Lin}
\email{sylin@cc.ncue.edu.tw}
\affiliation{Department of Physics, National Changhua University of Education, Changhua 50007, Taiwan}
\author{Kazutomu Shiokawa}
\email{kshiok@gmail.com}
\affiliation{Joint Quantum Institute, University of Maryland, College Park, Maryland 20742-4111, USA}
\author{Chung-Hsien Chou}
\email{chouch@mail.ncku.edu.tw}
\affiliation{Physics Division, National Center for Theoretical Sciences (South) and
Department of Physics, National Cheng Kung University, Tainan 701, Taiwan}
\author{B. L. Hu}
\email{blhu@umd.edu}
\affiliation{Joint Quantum Institute, University of Maryland, College Park, Maryland 20742-4111, USA}
\date{April 6, 2012}

\begin{abstract} 
We consider the quantum teleportation of continuous variables modeled by Unruh-DeWitt detectors coupled to a common quantum field initially in the Minkowski vacuum. An unknown coherent state of an Unruh-DeWitt detector is
teleported from one inertial agent (Alice) to an almost uniformly accelerated agent (Rob, for relativistic motion), using a detector pair initially entangled and shared by these two agents. The averaged physical fidelity of quantum teleportation, 
which is independent of the observer's frame, 
always drops below the best fidelity value from classical teleportation before the detector pair becomes disentangled
with the measure of entanglement evaluated around the future lightcone of the joint measurement event by Alice.
The distortion of the quantum state of the entangled detector pair from the initial state can suppress the fidelity
significantly even when the detectors are still strongly entangled around the lightcone.
We point out that the dynamics of entanglement of the detector pair observed in Minkowski frame or in quasi-Rindler
frame are not directly related to the physical fidelity of quantum teleportation in our setup.
These results are useful as a guide to making judicious choices of states and parameter ranges and estimation of the efficiency of quantum teleportation in relativistic quantum systems under environmental influences.
\end{abstract}


\maketitle

\section{Introduction}

Quantum teleportation is not only of practical values but also of theoretical interest because it contains many
illuminating manifestations of quantum physics, clarifying fundamental issues such as quantum information and classical
information, quantum nonlocality and relativistic locality, spacelike correlations 
and causality, and so on \cite{QTelepExpts}.
The first scheme of quantum teleportation is proposed by
Bennett, Brassard, Crepeau, Jozsa, Peres, and Wootters (BBCJPW) \cite{BBCJPW93},
where an {\it unknown} state of a qubit $C$ is teleported from one spatially localized agent Alice to another agent Bob using 
an entangled pair of qubits $A$ and $B$ prepared in one of the Bell states and shared by Alice and Bob, respectively. 
Such an idea is then adapted to the systems with continuous variables such as harmonic oscillators
by Vaidman \cite{Va94}, who introduces an EPR state \cite{EPR35} for the shared entangled pair to teleport an unknown coherent state.
Braunstein and Kimble (BK) \cite{BK98} generalized Vaidman's scheme from EPR states with exact correlations
to squeezed coherent states. In doing so the uncertainty of the measurable quantities
has to be considered, which reduces the degree of entanglement of the $AB$-pair
as well as the fidelity of teleportation.

Since quantum teleportation can address the issues of nonlocality and causality, it is natural to consider quantum
teleportation in fully relativistic systems and in non-inertial frames. Alsing and Milburn made the first attempt of 
calculating the fidelity of quantum teleportation between two moving cavities in relativistic motions \cite{AM03} -- 
one is at rest (Alice), the other is uniformly accelerated (Rob, for relativistic motion) in the Minkowski frame --
though their process of teleportation is not complete and the result is not quite reliable \cite{SU05, FM05}.
Alternatively, one of us \cite{Tom09} has considered quantum teleportation in the Unruh-DeWitt (UD) detector theory
\cite{Unr76, DeW79} with the agents in similar motions but based on the BK scheme:
An unknown coherent state of an UD detector $C$ is teleported from Alice to Rob using an entangled pair of the UD
detectors $A$ and $B$ initially in a two-mode squeezed state and held by Alice and Rob, respectively.
While both $A$ and $B$ are coupled with a common massless quantum field initially in the Minkowski vacuum,
the detector $C$ is isolated from the environment to identify the best fidelity that $A$ and $B$ moving
in a common environment can offer.

Unfortunately, the fidelity of quantum teleportation considered in \cite{Tom09} is not a physical one. 
More careful consideration of the relativistic effects of quantum information associated with the quantum field 
is needed to get the correct results, which we do in this paper.

There are many subtle issues related to quantum teleportation in relativistic systems which play an important role but have 
hitherto been largely ignored. Foremost, it is crucial that Bob or Rob has to have the full knowledge about the quantum state 
of the $AB$-pair to achieve quantum teleportation perfectly. For example, in the BBCJPW scheme,
Alice performs a Bell measurement jointly on $C$ and $A$, which collapses the $CA$-pair
into one of the Bell states, then sends the information to Bob classically about which Bell state the $CA$-pair ends up in.
According to this classical information Bob can reproduce the initial state of $C$ by operating on $B$. Bob's ``Manual" in
how to connect each kind of received information (the outcome of measurement on the $CA$-pair) with some specific set of
operations on $B$ is determined by the quantum state of the $AB$-pair right before the joint measurement by Alice.
If the quantum state of the entangled $AB$-pair is changed from their initial state, then Bob's ``Manual" must be changed
accordingly to keep the fidelity of quantum teleportation perfect.

Second, categorically there seems to be  a common misunderstanding or underestimation of {\it relativistic effects}
in quantum information. Many people may think of ``relativistic" as referring to fast motion of the atoms (qubits) or oscillators
(detectors), which is not commonly encountered, but forget about the fact that a quantum field which is ever present in any setup
is relativistic, namely that the laws of special relativity govern the field and hence enter in the interaction between the qubits
or detectors even when there is no direct interaction amongst them. This oversight is largely due to the focus of quantum foundational
issues from the viewpoint of quantum mechanics, which is only an approximation -- the nonrelativistic limit --
to the consistent 
theory, which is quantum field theory.

Third, which is related to the second, is the neglect of {\it  environmental influences}. The presence of a quantum field
is unavoidable and acts as an ubiquitous environment to the qubits or detectors in question ($A$, $B$, or $C$ described in the context
of quantum teleportation above) whose effects need to be included in one's consideration. They include:
\paragraph{Quantum decoherence}
Each qubit or oscillator can be decohered by coupling to a quantum field, but decoherence can be lessened if the two qubits are
placed in close range due to mutual influences mediated by the field; 
\paragraph{Entanglement dynamics}
The entanglement between two qubits or oscillators changes in time as their reduced state evolves; \
\paragraph{Unruh effect}
A pointlike object such as a UD detector coupled with a quantum field and uniformly accelerated in the Minkowski vacuum
of that quantum field would experience a thermal bath of the field quanta at the Unruh temperature proportional to
its proper acceleration.

Furthermore, objects in a relativistic system may behave differently when observed in different reference frames:
\paragraph{Frame dependence}
Since quantum entanglement between two spatially localized degrees of freedom is a kind of spacelike correlation in a
quantum state, which depends on reference frames, quantum entanglement of two localized objects but separated
in space is frame-dependent;
\paragraph{Time dilation}
For two moving objects localized in space and so parameterized by their proper times, the time dilations
of them observed in the rest frame will naturally enter the dynamics of entanglement between them.

The above factors have been considered in some detail \cite{ASH, LH09, LCH08}.
But there are new issues of foundational value which need be included in the consideration of quantum teleportation.
We mention two such issues related to open quantum systems below.

\subsection{Fidelity and entanglement in open quantum systems}

As indicated in the BK scheme, fidelity of quantum teleportation could depend on (i) quantum entanglement of the $AB$-pair,
and (ii) the consistency of the quantum state of the $AB$-pair with their initial state. Both would be reduced by the coupling
with an environment, the quantum field being an ubiquitous one. We will 
compare the time evolution of the logarithmic negativity of the $AB$-pair in a given reference frame
and the fidelities of quantum teleportation in the same frame to distinguish the effects caused by these two factors.

For simplicity and for the purpose of comparing the degree of entanglement of the $AB$-pair 
and the fidelity of quantum teleportation,
one of us \cite{Tom09} considered the ``pseudo-fidelity" of teleportation by imagining that right at the moment the
joint measurement on the $CA$-pair was done by Alice, Rob gets the information of the outcome from Alice instantaneously
and performs proper local operations on $B$ accordingly. In reality classical information need some time to travel from Alice to Rob,
and during the traveling time of the signal detector $B$ keeps evolving, so we expect the physical fidelity of teleportation
will be further reduced. Our calculation verifies this feature.

\subsection{Measurement in different frames}
Quantum states make sense only in a given frame where a Hamiltonian is well defined \cite{AA81}.
Two quantum states of the same system with quantum fields in different frames are comparable only on those totally overlapping
time-slices associated with certain moments in each frame. By a measurement local in space, e.g. on a point-like UD detector
coupled with a quantum field, quantum states (of the combined system) in different frames can be interpreted as if they
collapsed on different time-slices passing through the same measurement event. Nevertheless, the post-measurement states will
evolve to the same state up to a coordinate transformation when they are compared at some time-slice in the future \cite{Lin11a}.

Given the consistency of instantaneous measurement, we can further study the evolution of fidelities of teleportation in
different frames. Although they don't have to agree with each other if the time-slices of those different frames never overlap
(except for the fiducial time-slice where the initial state is defined), we will show that the reduced state of the qubit or
detector $B$ after wave functional collapse in different frames will become consistent once Rob enters the future lightcone of
the joint measurement event by Alice. To calculate the physical fidelity, we find it the most convenient to collapse the wave
functional of the combined system almost on the future lightcone of the local measurement event by Alice so the continuous evolution during the
moment of wave functional collapse and the local operation by Rob can be neglected in the cases with negligible mutual influences.

The paper is organized as follows. In Section II 
we introduce the model we use for addressing these issues. 
In Section III we review the definition of the averaged pseudo-fidelity 
and calculate the pseudo-fidelities in the ultraweak coupling limit in different frames; we
then illustrate some results beyond the ultraweak coupling limit.
In Section IV we modify the setup and calculate the averaged physical fidelity for a more realistic case
in the ultraweak coupling limit. We then summarize our findings in Section V.
In Appendix A we show explicitly that quantum entanglement of the $AB$-pair is a
necessary condition for the best fidelity of quantum teleportation beating the best classical fidelity
in the ultraweak coupling limit of our model.
Finally a short discussion on nonlocality and causality is given in Appendix B.

\section{The Model}
\label{model}

To address the above issues, we consider three identical Unruh-DeWitt detectors $A$, $B$, and $C$ with mass
$m=1$ and natural frequency $\Omega$, moving in a quantum field in (3+1)D Minkowski space.
The action of the combined system is given by \cite{LCH08}
\begin{eqnarray}
  S &=& -\int d^4 x \sqrt{-g} {1\over 2}\partial_\mu\Phi(x) \partial^\mu\Phi(x) +\nonumber\\ & &
    \sum_{{\bf d}=A,B,C}\int d\tau_{\bf d} \left\{ {1\over 2}\left[\left(\partial_{\bf d}Q_{\bf d}\right)^2
    -\Omega_{0}^2 Q_{\bf d}^2\right]
    + \lambda_0\int d^4 x 
    Q_{\bf d}(\tau_{\bf d})\Phi (x)\delta^4\left(x^{\mu}-z_{\bf d}^{\mu}(\tau_{\bf d})\right)\right\},
  \label{Stot1}
\end{eqnarray}
where $\mu, \nu=0,1,2,3$, $g_{\mu\nu} = {\rm diag}(-1,1,1,1)$, $\partial^{}_{\bf d}\equiv \partial/\partial \tau^{}_{\bf d}$,
$\tau^{}_A$, $\tau^{}_B$ and $\tau^{}_C$ are proper times for $Q_A$, $Q_B$, and $Q_C$, respectively. The
scalar field $\Phi$ is assumed to be massless, and $\lambda_0$ is the coupling constant.
Detectors $A$ and $C$ are held by Alice, who is at rest in space with the worldline
$z_A^\mu =z_C^\mu = (t, 1/b,0,0)$, while $B$ is held by Rob, who is uniformly accelerated along the worldline
$z_B^\mu = (a^{-1}\sinh a\tau, a^{-1}\cosh a\tau,0,0)$, $0<a<b$, where $\tau$ is Rob's proper time,
namely, $\tau^{}_A=\tau^{}_C=t$ and $\tau^{}_B=\tau$.

Suppose the initial state of the combined system at $t=\tau=0$ is a product state $\rho^{}_{\Phi_{\bf x}}\otimes\rho^{}_{AB}
\otimes\rho_C^{(\alpha)}$ of the Minkowski vacuum of the field $\hat{\rho}^{}_{\Phi_{\bf x}} = \left| 0_M\right>\left< 0_M\right|$,
a two-mode squeezed state of the detectors $A$ and $B$, in the $(K, \Delta)$ representation \cite{UZ89}
(or the ``Wigner characteristic function" \cite{GZ99}),
\begin{eqnarray}
  & & \rho^{}_{AB}(K^A, K^B, \Delta^A, \Delta^B) = \nonumber\\ & & 
  \exp -{1\over 2\hbar}\left[ {e^{2r_1}\over\Omega}(K^A+K^B)^2 + \Omega e^{-2r_1} (\Delta^A +\Delta^B)^2 +
  {e^{-2r_1}\over\Omega} (K^A - K^B)^2+ \Omega e^{2r_1} (\Delta^A -\Delta^B)^2 \right],
\label{rhoABI}
\end{eqnarray}
and a coherent state of the detector $C$, denoted $\hat{\rho}_C^{(\alpha)} =
\left|\alpha \right>^{}_C\left<\alpha\right|$, or in the $(K, \Delta)$ representation $(\alpha=\alpha^{}_R + i\alpha^{}_I)$,
\begin{equation}
   \rho_C^{(\alpha)}(K^C, \Delta^C) = \exp \left[ -{1\over 2\hbar}\left( {1\over 2\Omega} (K^C)^2+
   {\Omega\over 2}(\Delta^C)^2 \right)+{i\over \hbar} \left(
  \sqrt{2\hbar\over\Omega} \alpha^{}_R K^C - \sqrt{2\hbar\Omega}\alpha^{}_I \Delta^C \right) \right].
\label{rhoAl}
\end{equation}
$\rho_C^{(\alpha)}$ is the quantum state to be teleported. In general the factors in $\rho_C^{(\alpha)}$
will vary in time. To concentrate on the best fidelity of quantum teleportation that the entangled $AB$-pair can offer, however,
we follow Ref. \cite{Tom09} and assume the dynamics of $\rho_C^{(\alpha)}$ is frozen, or equivalently, assume $\rho_C^{(\alpha)}$
is created just before teleportation. Note that, as $r_1\to \infty$, $\rho^{}_{AB}$ goes to an EPR state with the correlations
$\left<\right. Q_A-Q_B \left.\right>= \left<\right. P_A+P_B \left.\right>=0$ without uncertainty,
while $Q_A+Q_B$ and $P_A-P_B$ are totally uncertain.

At $t=\tau=0$ in the Minkowski frame, the detectors $A$ and $B$ start to couple with the field,
while the detector $C$ is isolated from others.
By virtue of the linearity of the combined system $(\ref{Stot1})$, operators at some coordinate time $x^0$ after the initial moment
are linear combinations of the operators defined at the initial moment \cite{LH06}:
\begin{eqnarray}
  \hat{Q}^{}_{\bf d}(\tau^{}_{\bf d}(x^0)) &=& \sum_{{\bf d}'}
    \left[\phi^{{\bf d}'}_{\bf d}(\tau^{}_{\bf d})\hat{Q}^{[0]}_{{\bf d}'} +
    f^{{\bf d}'}_{\bf d}(\tau^{}_{\bf d})\hat{P}^{[0]}_{{\bf d}'} \right]+
    \int d^3y \left[ \phi^{\bf y}_{\bf d}(\tau^{}_{\bf d})\hat{\Phi}^{[0]}_{\bf y} +
    f^{\bf y}_{\bf d}(\tau^{}_{\bf d})\hat{\Pi}^{[0]}_{\bf y} \right], \label{Qexp}\\
  \hat{\Phi}^{}_{\bf x}(x^0) &=& \sum_{{\bf d}'} \left[\phi^{{\bf d}'}_{\bf x}(x^0)\hat{Q}^{[0]}_{{\bf d}'} +
    f^{{\bf d}'}_{\bf x}(x^0)\hat{P}^{[0]}_{{\bf d}'}\right] + \int d^3y \left[
    \phi^{\bf y}_{\bf x}(x^0)\hat{\Phi}^{[0]}_{\bf y} + f^{\bf y}_{\bf x}(x^0)\hat{\Pi}^{[0]}_{\bf y} \right], \label{Phiexp}
\end{eqnarray}
from which $\hat{P}^{}_{\bf d}(x^0)$ and $\hat{\Pi}^{}_{\bf x}(x^0)$ can be derived.
Here $\hat{\cal O}_{\zeta}^{[n]} \equiv \hat{\cal O}_{\zeta}(t_n)$ and
all the ``mode functions" $\phi^{\zeta}_\xi(x^0)$ and $f^{\zeta}_\xi(x^0)$ are real functions of time
($\zeta, \xi = \{{\bf d}\}\cup \{{\bf x}\}$),
which can be related to those in $k$-space in Ref. \cite{LH06}.

Comparing the expansions $(\ref{Qexp})$ and $(\ref{Phiexp})$ of two equivalent continuous evolutions, one from
$x^0_{\bf 0}\equiv 0$ to $x^0_{\bf 1}$ then from $x^0_{\bf 1}$ to $x^0_{\bf 2}$, the other from $x^0_{\bf 0}$ all the way to
$x^0_{\bf 2}$, one can see that the mode functions have the following identities,
\begin{eqnarray}
  & &\phi^{\zeta[20]}_\xi  = \sum_{{\bf d}'}\left[\phi_\xi^{{\bf d}'[21]}\phi_{{\bf d}'}^{\zeta[10]} +
    f_\xi^{{\bf d}'[21]} \pi^{\zeta[10]}_{{\bf d}'}\right] + 
  \int d^3x' \left[ \phi^{{\bf x'}[21]}_\xi \phi^{\zeta[10]}_{\bf x'} + f^{{\bf x'}[21]}_\xi \pi^{\zeta[10]}_{\bf x'}\right], \label{id1}\\
  & & f^{\zeta[20]}_\xi  = \sum_{{\bf d}'}\left[\phi_\xi^{{\bf d}'[21]}f_{{\bf d}'}^{\zeta[10]} +
    f_\xi^{{\bf d}'[21]} p^{\zeta[10]}_{{\bf d}'}\right] + 
  \int d^3x' \left[ \phi^{{\bf x'}[21]}_\xi f^{\zeta[10]}_{\bf x'} + f^{{\bf x'}[21]}_\xi p^{\zeta[10]}_{\bf x'}\right], \label{id2}
\end{eqnarray}
where $F^{[mn]} \equiv F(x^0_m-x^0_n)$,
$\pi^{\zeta}_{\bf d}(\tau^{}_{\bf d}(x^0)) \equiv \partial^{}_{\bf d}\phi^{\zeta}_{{\bf d}}(\tau^{}_{\bf d}(x^0))$,
$\pi^{\zeta}_{\bf x}(x^0) \equiv \partial_0 \phi^{\zeta}_{\bf x}(x^0)$,
$p^{\zeta}_{\bf d}(\tau^{}_{\bf d}(x^0)) \equiv \partial^{}_{\bf d}f^{\zeta}_{{\bf d}}(\tau^{}_{\bf d}(x^0))$,
and $p^{\zeta}_{\bf x}(x^0) \equiv \partial_0 f^{\zeta}_{\bf x}(x^0)$.
Similar identities for $\pi^\zeta_\xi$ and $p^\zeta_\xi$ can be derived straightforwardly from $(\ref{id1})$ and $(\ref{id2})$.
Such identities can be interpreted as embodying the Huygens' principle of the mode functions,
and can be verified by inserting particular solutions of the mode functions into the identities.

By virtue of the linearity of the combined system $(\ref{Stot1})$, the quantum state of the combined system started
with a Gaussian state will always be Gaussian;  therefore the reduced state of the three detectors is Gaussian for all times.
In the $(K,\Delta)$ representation the reduced Wigner function at the coordinate time $x^0$ in the reference frame
of some observer has the form
\begin{eqnarray}
  \rho^{}_{ABC}({\bf K},{\bf \Delta};x^0) &=& \exp\left[ -{1\over 2\hbar^2} \left( K^{\bf d} {\cal Q}_{{\bf d}{\bf d'}}(x^0) K^{\bf d'}
  +\Delta^{\bf d} {\cal P}_{{\bf d}{\bf d'}}(x^0) \Delta^{\bf d'} - 2 K^{\bf d} {\cal R}_{{\bf d}{\bf d'}}(x^0) \Delta^{\bf d'}\right)
  \right.\nonumber\\  & & \left.+ {i\over \hbar} \left( \left<\right. \hat{Q}_{\bf d}(x^0)\left.\right> K^{\bf d} -
  \left<\right. \hat{P}_{\bf d}(x^0)\left.\right> \Delta^{\bf d}\right) \right],
\label{rhoABC}
\end{eqnarray}
where ${\bf d}, {\bf d'} = A,B,C$, and the factors
\begin{eqnarray}
{\cal Q}_{{\bf d}{\bf d'}}(x^0) &=& {\hbar\delta\over i\delta K^{\bf d}} {\hbar\delta\over i\delta K^{\bf d'}}
  \rho^{}_{ABC}({\bf K},{\bf \Delta};x^0)|_{{\bf K}={\bf \Delta}=0} =
  \left<\right. \delta Q_{\bf d}(\tau_{\bf d}(x^0)), \delta Q_{\bf d'}(\tau_{\bf d'}(x^0)) \left.\right>,\\
{\cal P}_{{\bf d}{\bf d'}}(x^0) &=& {i\hbar\delta\over \delta \Delta^{\bf d}} {i\hbar\delta\over \delta \Delta^{\bf d'}}
  \rho^{}_{ABC}({\bf K},{\bf \Delta};x^0)|_{{\bf K}={\bf \Delta}=0} =
  \left<\right. \delta P_{\bf d}(\tau_{\bf d}(x^0)), \delta P_{\bf d'}(\tau_{\bf d'}(x^0)) \left.\right>,\\
{\cal R}_{{\bf d}{\bf d'}}(x^0) &=& {\hbar\delta\over i\delta K^{\bf d}} {i\hbar\delta\over \delta \Delta^{\bf d'}}
  \rho^{}_{ABC}({\bf K},{\bf \Delta};x^0)|_{{\bf K}={\bf \Delta}=0} =
  \left<\right. \delta Q_{\bf d}(\tau_{\bf d}(x^0)), \delta P_{\bf d'}(\tau_{\bf d'}(x^0)) \left.\right>,
\end{eqnarray}
are actually those symmetric two-point correlators of the detectors in their covariance matrices
($\left<\right. {\cal O}, {\cal O}'\left.\right>\equiv \left<\right. {\cal O}{\cal O}'+{\cal O}'{\cal O}\left.\right>/2$
and $\delta {\cal O} \equiv \hat{\cal O}-\left<\right.\hat{\cal O}\left.\right>$),
which can be obtained in the Heisenberg picture by taking the expectation values of
the evolving operators with respect to the initial state.
From $(\ref{Qexp})$ and $(\ref{Phiexp})$, these correlators are combinations of
the mode functions and the initial data, e.g.,
\begin{equation}
  \left<\right. \hat{Q}_A^2(\tau^{}_A)\left.\right> =
  \phi_A^A(\tau^{}_A)\phi_A^A(\tau^{}_A)\left<\right. (\hat{Q}_A^{[0]})^2\left.\right>_0 +
   \int d^3x d^3y\, \phi_A^{\bf x}(\tau^{}_A)\phi_A^{\bf y}(\tau^{}_A)\left<\right. \hat{\Phi}_{\bf x}^{[0]},\hat{\Phi}_{\bf y}^{[0]}
   \left.\right>_0 + \ldots,
\label{QA2examp}
\end{equation}
where $\left<\right. \cdots \left.\right>_n$ denotes that the expectation values are taken from the quantum state right after $t=t_n$
($t_0=0$ here.)

Suppose the reduced state of the three detectors continuously evolve to $\rho^{}_{ABC}({\bf K},{\bf \Delta};t_1)$ in the Minkowski frame
at some moment $t=t_1>0$ and $\tau=\tau_1 \equiv a^{-1}\sinh^{-1} at_1$, when a joint Gaussian measurement by Alice is performed
{\it locally in space} on $A$ and $C$ so that the post-measurement state right after $t_1$ in the Minkowski frame becomes
$\tilde{\rho}^{}_{ABC}({\bf K},{\bf \Delta};t_1)= \tilde{\rho}^{(\beta)}_{AC}(K^A, K^C,\Delta^A, \Delta^C)\tilde{\rho}_{B}(K^B, \Delta^B)$,
where we assume the quantum state of detectors $A$ and $C$ becomes another two-mode squeezed state
\footnote{The state $(\ref{PMSAC})$ is chosen so that the analytic calculation is the simplest while the result is still interesting.
One may choose another state consistent with the EPR state as the squeeze parameter $r_2\to\infty$ instead,
for example, $K^C$ and $\Delta^C$ are replaced by $(K^C-K^C)$ and $(\Delta^C+\Delta^A)$,
respectively. Then $N_B$ and the $F_{av}$ will be more complicated and will depend on $\alpha$.
In practice the choice of the state may depend on the experimental setting.}
\begin{eqnarray}
  & &\tilde{\rho}^{(\beta)}_{AC}(K^A, K^C,\Delta^A, \Delta^C) = \nonumber\\ & & \exp\left[ -{1\over 2\hbar^2}
  \left( K^m \tilde{\cal Q}_{mn} K^n +\Delta^m \tilde{\cal P}_{mn} \Delta^n - 2 K^m \tilde{\cal R}_{mn} \Delta^n\right) +
  {i\over \hbar} \left( \sqrt{2\hbar\over\Omega}\beta_R K^C -\sqrt{2\hbar\Omega}\beta_I\Delta^C\right) \right],
\label{PMSAC}
\end{eqnarray}
with $m,n = A,C$ so that Alice gets the outcome $\beta = \beta_R + i\beta_I$.
Then $(\ref{PMSAC})$ yields the
reduced state of detector $B$
\begin{eqnarray}
  \tilde{\rho}^{}_B(K^B,\Delta^B) &=& N_B \int {dK^C d\Delta^C \over 2\pi\hbar}{dK^A d\Delta^A \over 2\pi\hbar}
  \tilde{\rho}_{AC}^{(\beta) *}(K^A, K^C, \Delta^A,\Delta^C)\rho^{}_{ABC}(K^A, K^B, K^C,\Delta^A,\Delta^B,\Delta^C; t_1),
\label{rhoB}
\end{eqnarray}
where $N_B$ is the normalization constant. If we require ${\rm Tr}^{}_B\, \tilde{\rho}^{}_B= \tilde{\rho}^{}_B|_{K^B=\Delta^B=0}=1$,
then $N_B$ will depend on $\beta$. Alternatively, following \cite{Tom09}, we can require $N_B$ to be independent
of $\beta$, then ${\rm Tr}^{}_B\, \tilde{\rho}^{}_B$ will be proportional to the probability $P(\beta)$ of finding detectors $A$ and $C$
in the state $(\ref{PMSAC})$. Let ${\rm Tr}^{}_B\, \tilde{\rho}^{}_B = P(\beta)$, then we have the normalization condition
\begin{eqnarray}
  1&=&\int d^2\beta P(\beta)= \int d\beta^{}_R d\beta^{}_I \, \tilde{\rho}^{}_B(K^B=0,\Delta^B=0)\nonumber\\
  &=& N_B\int d\beta_R d\beta_I {dK^A d\Delta^A\over 2\pi \hbar}{dK^C d\Delta^C\over 2\pi \hbar}
    \tilde{\rho}_{AC}^{(\beta) *}(K^A, K^C, \Delta^A,\Delta^C)\rho^{}_{ABC}( K^A,0, K^C, \Delta^A,0,\Delta^C; t_1) \nonumber\\
  &=& N_B \int {dK^A d\Delta^A\over 2\pi \hbar}{dK^C d\Delta^C\over 2\pi \hbar} \rho^{}_{ABC}(K^A,0, K^C, \Delta^A,0,\Delta^C; t_1)
    2\pi\delta\left(\sqrt{2\over\hbar\Omega}K^C\right)2\pi\delta\left(\sqrt{2\Omega\over\hbar}\Delta^C\right)\times
    \nonumber\\ & & \hspace{1cm}\exp \left[ -{1\over 2\hbar^2}
    \left( K^m \tilde{\cal Q}_{mn} K^n +\Delta^m \tilde{\cal P}_{mn} \Delta^n - 2 K^m \tilde{\cal R}_{mn} \Delta^n\right)\right]
    \nonumber\\
  &=& {N_B\over 2\hbar} \int dK^A d\Delta^A \exp {-1\over 2\hbar^2}
    \left[ \left({\cal Q}_{AA}^{[1]} + \tilde{\cal Q}_{AA}\right) (K^A)^2 +
    \left({\cal P}_{AA}^{[1]} + \tilde{\cal P}_{AA}\right) (\Delta^A)^2 -
    2 K^A \left({\cal R}_{AA}^{[1]} + \tilde{\cal R}_{AA}\right) \Delta^A\right],\nonumber
\end{eqnarray}
after inserting $(\ref{rhoABC})$ and $(\ref{PMSAC})$ into the integrand.
Here ${\cal S}^{[n]}$ denotes the value of the factor ${\cal S} = {\cal Q}, {\cal P}$, or ${\cal R}$ being taken at
$t_n-\epsilon$ with $\epsilon\to 0+$. Thus we have
\begin{equation}
  N_B = {1\over\pi\hbar}\sqrt{ \left({\cal Q}_{AA}^{[1]} + \tilde{\cal Q}_{AA}\right)
  \left({\cal P}_{AA}^{[1]} + \tilde{\cal P}_{AA}\right)- \left({\cal R}_{AA}^{[1]} + \tilde{\cal R}_{AA}\right)^2}.
\label{NormB}
\end{equation}

\section{Pseudo-Fidelities and Entanglement in Different Frames}

To compare the fidelity of quantum teleportation with quantum entanglement between detectors $A$ and $B$ at $t_1$,
which is defined on the same $t_1$-slice in the Minkowski frame, we first imagine that Rob receives the
outcome $\beta$ of Alice's joint measurement on $A$ and $C$ and make the proper operation on detector $B$
{\it instantaneously} at $\tau_1(t_1)$ when the worldline of $B$ intersects the $t_1$-slice (see Fig. \ref{AR}).
Physical situations with the classical signal from Alice traveling at the speed of light
will be considered later in Section \ref{physreal}.

In the BK scheme \cite{BK98, Tom09}, according to the outcome $\beta$ obtained by Alice,
the operation that Rob should perform on detector $B$ is a displacement by $\beta$ in phase space of $B$
from $\tilde{\rho}^{}_B$ to $\rho^{}_{out}$, namely, $\hat{\tilde{\rho}}^{}_{out} = \hat{D}(\beta)\hat{\rho}^{}_B$,
where $\hat{D}(\beta)$ is the displacement operator, or in the $(K,\Delta)$ representation,
\begin{equation}
  \rho_{out}(K^B, \Delta^B) =
  \tilde{\rho}^{}_B(K^B, \Delta^B) \exp {i\over\hbar}\left( \sqrt{2\hbar\over\Omega}\beta_R K^B
   -\sqrt{2\hbar\Omega}\beta_I \Delta^B\right).
\label{rhoOut}
\end{equation}
The ``pseudo-fidelity" of quantum teleportation from $|\alpha\left.\right>^{}_C$ to
$|\alpha\left.\right>^{}_B$ is then defined as
\begin{equation}
  F(\beta) \equiv {{}^{}_B\left< \right.\alpha\,|\hat{\rho}_{out}|\,\alpha\left.\right>^{}_B\over {\rm Tr}\hat{\rho}^{}_{out} }.
\end{equation}
Note that ${\rm Tr}^{}_B\hat{\rho}^{}_{out} = \rho^{}_{out}(K^B=0, \Delta^B=0) ={\rm Tr}^{}_B\hat{\rho}^{}_{B} = P(\beta)$.
A simpler quantity for calculation here is the {\it averaged} pseudo-fidelity, defined by
\begin{eqnarray}
  F_{av} &\equiv& \int d^2 \beta P(\beta) F(\beta) = \int d\beta_R d\beta_I \,
   {}^{}_B\left<\right.\alpha|\hat{\rho}_{out}|\alpha\left.\right>^{}_B \nonumber\\
  &=& \int d\beta_R d\beta_I {dK^B d\Delta^B\over 2\pi \hbar} \rho_B^{(\alpha) *}(K^B, \Delta^B) \rho_{out}(K^B,\Delta^B),
\end{eqnarray}
where $\hat{\rho}_B^{(\alpha)} =\left|\alpha \right>^{}_B\left<\alpha\right|$.
From $(\ref{rhoAl})$ and $(\ref{rhoOut})$, with the help of $(\ref{rhoB})$, $(\ref{rhoABC})$ and $(\ref{PMSAC})$, we have
\begin{eqnarray}
  F_{av} &=& N_B \int d\beta_R d\beta_I {\prod_{\bf d} dK^{\bf d} d\Delta^{\bf d} \over (2\pi \hbar)^3} \exp\left\{
    {i\over\hbar}\left[ \sqrt{2\hbar\over \Omega}(\alpha^{}_R-\beta^{}_R)(K^C-K^B)-
    \sqrt{2\hbar\Omega}(\alpha^{}_I-\beta^{}_I)(\Delta^C-\Delta^B)\right] +\right. \nonumber\\
    & &\left.-{1\over 2\hbar^2}\left[ {\hbar\over 2\Omega} (K^B)^2
    +{\hbar\over 2}\Omega (\Delta^B)^2 + K^m \tilde{\cal Q}_{mn} K^n +\Delta^m \tilde{\cal P}_{mn} \Delta^n - 2 K^m \tilde{\cal R}_{mn}
    \Delta^n\right] \right\} \rho^{}_{ABC}({\bf K}, {\bf \Delta}; t_1) \nonumber\\ &=&
  N_B \int {\prod_{\bf d} dK^{\bf d} d\Delta^{\bf d}\over (2\pi \hbar)^3} (2\pi)^2 \delta\left(\sqrt{2\over\hbar \Omega}(K^C-K^B)\right)
    \delta\left(\sqrt{2\Omega \over\hbar}(\Delta^C-\Delta^B)\right) \rho^{}_{ABC}({\bf K}, {\bf \Delta}; t_1)\times \nonumber\\ & &
    \exp \left\{-{1\over 2\hbar^2}\left[ {\hbar\over 2\Omega} (K^B)^2
    +{\hbar\over 2}\Omega (\Delta^B)^2 + K^m \tilde{\cal Q}_{mn} K^n +\Delta^m \tilde{\cal P}_{mn} \Delta^n - 2 K^m \tilde{\cal R}_{mn}
    \Delta^n\right] \right\},
\end{eqnarray}
thus
\begin{equation}
  F_{av}  = {\hbar^2\pi N_B \over \sqrt{\det \tilde{\bf V}}} ,
\label{Favformula}
\end{equation}
where
\begin{equation}
 \tilde{\bf V} = \left(\begin{array}{cccc}
 {\cal Q}_{AA}^{[1]}+\tilde{\cal Q}_{AA} & -{\cal R}_{AA}^{[1]}-\tilde{\cal R}_{AA} &
 {\cal Q}_{AB}^{[1]}+\tilde{\cal Q}_{AC} &  -{\cal R}_{AB}^{[1]}-\tilde{\cal R}_{AC} \\
 -{\cal R}_{AA}^{[1]}-\tilde{\cal R}_{AA} & {\cal P}_{AA}^{[1]}+\tilde{\cal P}_{AA} &
 -{\cal R}_{BA}^{[1]}-\tilde{\cal R}_{CA} &  {\cal P}_{AB}^{[1]}+\tilde{\cal P}_{AC} \\
 {\cal Q}_{AB}^{[1]}+\tilde{\cal Q}_{AC} & -{\cal R}_{BA}^{[1]}-\tilde{\cal R}_{CA} &
 {\cal Q}_{BB}^{[1]}+\tilde{\cal Q}_{CC}+ \hbar \Omega^{-1} & -{\cal R}_{BB}^{[1]}-\tilde{\cal R}_{CC} \\
 -{\cal R}_{AB}^{[1]}-\tilde{\cal R}_{AC} & {\cal P}_{AB}^{[1]}+\tilde{\cal P}_{AC} &
 -{\cal R}_{BB}^{[1]}-\tilde{\cal R}_{CC} & {\cal P}_{BB}^{[1]}+\tilde{\cal P}_{CC}+\hbar\Omega
 \end{array}\right).
\label{tildeV}
\end{equation}
Note that $F_{av}$ in $(\ref{Favformula})$ is independent of $\alpha$ because of the choice of the state $(\ref{PMSAC})$.

Below we consider the cases with the factors in the two-mode squeezed state $(\ref{PMSAC})$ of detectors $A$ and $C$ right
after the joint measurement given by:  $\tilde{\cal Q}_{AA}=\tilde{\cal Q}_{CC}= {\hbar\over 2\Omega}\cosh 2r_2$, $\tilde{\cal Q}_{AC}=
{\hbar\over 2\Omega}\sinh 2r_2$, $\tilde{\cal P}_{AA}=\tilde{\cal P}_{CC}={\hbar\over 2}\Omega\cosh 2 r_2$,
$\tilde{\cal P}_{AC}= -{\hbar\over 2}\Omega\sinh 2r_2$ with squeezed parameter $r_2$, and $\tilde{\cal R}_{mn}=0$.

If the joint measurement on detectors $A$ and $C$ is done perfectly such that $r_2\to \infty$, then
from $(\ref{Favformula})$, $(\ref{tildeV})$, and $(\ref{NormB})$, we have
\begin{equation}
  F_{av}(t_1,\tau_1) \to \left[ \left( {1\over\Omega}+{1\over \hbar}\left<\right.Q_-^2\left.\right>\right)
  \left(\Omega+{1\over\hbar}\left<\right.P_+^2\left.\right>\right)
  -\left(\left<\right.Q_-,P_+\left.\right>\right)^2\right]^{-1/2},
\end{equation}
where $Q_- \equiv Q_A(t_1)-Q_B(\tau_1)$ and $P_+ \equiv P_A(t_1)+P_B(\tau_1)$.
For $t_1 = t_0=0$, the initial state $\rho^{}_{AB}$ of detectors $A$ and $B$ in $(\ref{rhoABI})$ without coupling
to the field gives
\begin{equation}
  F_{av}(0,0) = {1\over 1+e^{-2r_1}},
\end{equation}
which implies $F_{av}\to 1$ as $r_1 \to \infty$ when $\rho^{}_{AB}$ is nearly an EPR state, while
$F_{av}\to 1/2$ as $r_1 \to 0$ when $\rho^{}_{AB}$ is almost the coherent state of free detectors.
In the latter case $F_{av}=1/2$ is known as the best fidelity of ``classical" teleportation
using coherent states \cite{BK98},
without considering the coupling of the UD detectors with the environment.
This does not imply that $F_{av}$ of quantum teleportation must be greater than $1/2$.
In our result, if we start with the state with $r_1=0$, then $F_{av}$ will always be less than $1/2$ after the detectors
are coupled to the field, and detectors $A$ and $B$ are always separable, too.
Once the correlations such as $\left<Q_-\right>=0$ needed in the protocol of quantum teleportation becomes
more uncertain than the minimum quantum uncertainty, $F_{av}-1/2$ will become negative.

\subsection{ultraweak coupling limit}
\label{infiweak}

In the ultraweak coupling limit, $\gamma$ is so small that $\gamma\Lambda_1 \ll a, \Omega$.
Inserting the expressions for the correlators in this limit \cite{LCH08}, one obtains
$(\ref{QAAwc})$-$(\ref{RAAwc})$ and
\begin{equation}
  \tilde{\bf V} \approx
  \left( \begin{array}{cccc}
   {\hbar\over 2\Omega} {\cal A}(t_1) & 0 & {\hbar\over 2\Omega}  {\cal X}(t_1,\tau_1)& {\hbar\over 2} {\cal Y}(t_1,\tau_1) \\
  0 & {\hbar\over 2} \Omega {\cal A}(t_1)+\upsilon &
    {\hbar\over 2} {\cal Y}(t_1,\tau_1) & -{\hbar\over 2}\Omega {\cal X}(t_1,\tau_1)\\ 	
  {\hbar\over 2\Omega} {\cal X}(t_1,\tau_1) & {\hbar\over 2} {\cal Y}(t_1,\tau_1) & {\hbar\over 2\Omega} {\cal B}(\tau_1) & 0 \\
  {\hbar\over 2} {\cal Y}(t_1,\tau_1) & -{\hbar\over 2}\Omega {\cal X}(t_1,\tau_1) & 0 &
    {\hbar\over 2} \Omega {\cal B}(\tau_1)+\upsilon
\end{array}\right)
\label{tildeVwc}
\end{equation}
by writing $\upsilon\equiv 2\hbar\gamma\Lambda_1/\pi$.
Here $t^{}_1(x_{\bf 1}^0)$ and $\tau^{}_1(x_{\bf 1}^0)$ are the proper times of
detectors $A$ and $B$, respectively, when Alice performs the joint measurement on detectors $A$ and $C$ at coordinate time
$x_{\bf 1}^0$ observed in some reference frame, and
\begin{eqnarray}
  {\cal A}(t_1) &\equiv& C_2 +e^{-2\gamma t_1} C_1  +1-e^{-2\gamma t_1},\label{Aoft}\\
  {\cal B}(\tau_1) &\equiv& 2+C_2 + e^{-2\gamma\tau_1}C_1  + \left(1-e^{-2\gamma\tau_1}\right)\coth{\pi\Omega\over a},\label{BofT}\\
  {\cal X}(t_1,\tau_1) &\equiv& S_2 + e^{-\gamma(t_1+\tau_1)} \cos\Omega(t_1+\tau_1)\, S_1,\\
  {\cal Y}(t_1,\tau_1) &\equiv& e^{-\gamma(t_1+\tau_1)}\sin\Omega(t_1+\tau_1)\,S_1 ,
\end{eqnarray}
with $C_n \equiv \cosh 2r_n$ and $S_n\equiv \sinh 2r_n$. So the averaged pseudo-fidelity in the ultraweak coupling limit can be
written in a simple form,
\begin{equation}
  F_{av}(t_1, \tau_1) = {2 {\cal A}\over {\cal AB}-({\cal X}^2+{\cal Y}^2)} + O(\upsilon), \label{weakFav}
\end{equation}
where
\begin{equation}
  {\cal X}^2+{\cal Y}^2 = S_2^2+ S_1^2\, e^{-2\gamma(t_1+\tau_1)}+2S_1 S_2 \, e^{-\gamma(t_1+\tau_1)}\cos\Omega(t_1+\tau_1)
\label{X2Y2}
\end{equation}
is oscillating in time due to the natural squeeze-antisqueeze oscillation of the two-mode squeezed state of detectors $A$ and $B$.
The maximum (minimum) values of $F_{av}$, denoted by $F^+_{av}$ ($F^-_{av}$), occur at
$\cos\Omega(t_1+\tau_1)\approx 1$ ($-1$), when ${\cal Y}=0$ and
\begin{equation}
  F_{av}^{\pm}(t_1, \tau_1) \approx {2{\cal A}\over {\cal AB} - \left[S_2\pm S_1 \, e^{-\gamma(t_1+\tau_1)}\right]^2}.
\label{maxminFav}
\end{equation}

In the BBCJPW scheme, Alice and Rob have full knowledge of the entangled AB-pair. In the BK scheme for continuous variables, we may assume
the same thing: while the two-mode squeezed state of $A$ and $B$ squeezes and antisqueezes alternatingly in time, Alice completely knows
when the AB-pair will give the best correlation needed and thus the best fidelity of quantum teleportation with peak values
$F_{av}^+$, so she always performs the joint measurement on $A$ and $C$ at one of those moments. This actually requires that Alice
has had the full knowledge about how Rob moves to guarantee that $\tau(t_1)$ would make $\cos\Omega(t_1+\tau(t_1))\approx 1$.

In Appendix A we show that, in the ultraweak coupling limit of our model with the initial state $(\ref{rhoABI})$ and the
post-measurement state $(\ref{PMSAC})$, whenever detectors $A$ and $B$ are separable in some frame, the averaged
pseudo-fidelity of quantum teleportation $F_{av}$ must have been less than the best classical fidelity $1/2$ in that frame. In other words,
quantum entanglement between detectors $A$ and $B$ is necessary to provide the advantage of quantum teleportation,
at least in the ultraweak coupling limit of our model.

\subsubsection{Minkowski frame}

\begin{figure}
\includegraphics[width=7.5cm]{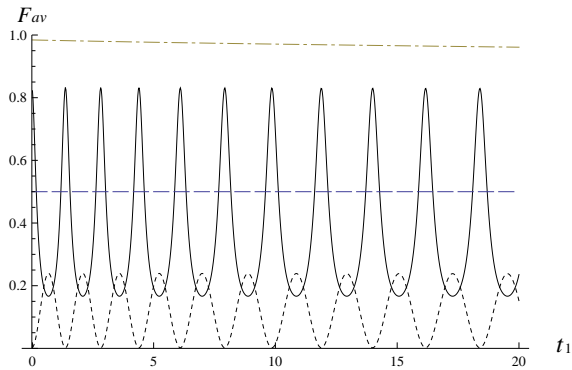}
\caption{A comparison of the averaged pseudo-fidelity $F_{av}(t_1)$ (solid curve) in the Minkowski frame in the ultraweak
coupling limit, the evolution of the correlator $\left<\right.Q_-^2\left.\right>/20$ (dotted curve), $Q_-\equiv Q_A - Q_B$
and the logarithmic negativity $E_{\cal N}/3.5$ (dot-dashed curve) at early times. One can see that $F_{av}$ reaches a maximum whenever
$\left<\right.Q_-^2\left.\right>$ reaches a minimum, while the logarithmic negativity $E_{\cal N}$  evolves smoothly remaining always well above zero during the same time interval.
So we can see that the oscillation of $F_{av}$ is due to the natural oscillation of the initial two-mode squeezed state,
and what causes $F_{av}$ to drop below $1/2$ here comes from the distortion of the quantum states of detectors $A$ and $B$ from their initial state; it is not an indication of disentanglement of the detectors. Here $\gamma = 0.0002$, $\Omega = 2.3$, $m=\hbar=1$, $r_1=1.1$, $r_2=1.2$,
$\Lambda_0=\Lambda_1=20$, $a=1/4$, and $b=2.01 a$.}
\label{wcCorr}
\end{figure}

\begin{figure}
\includegraphics[width=7.5cm]{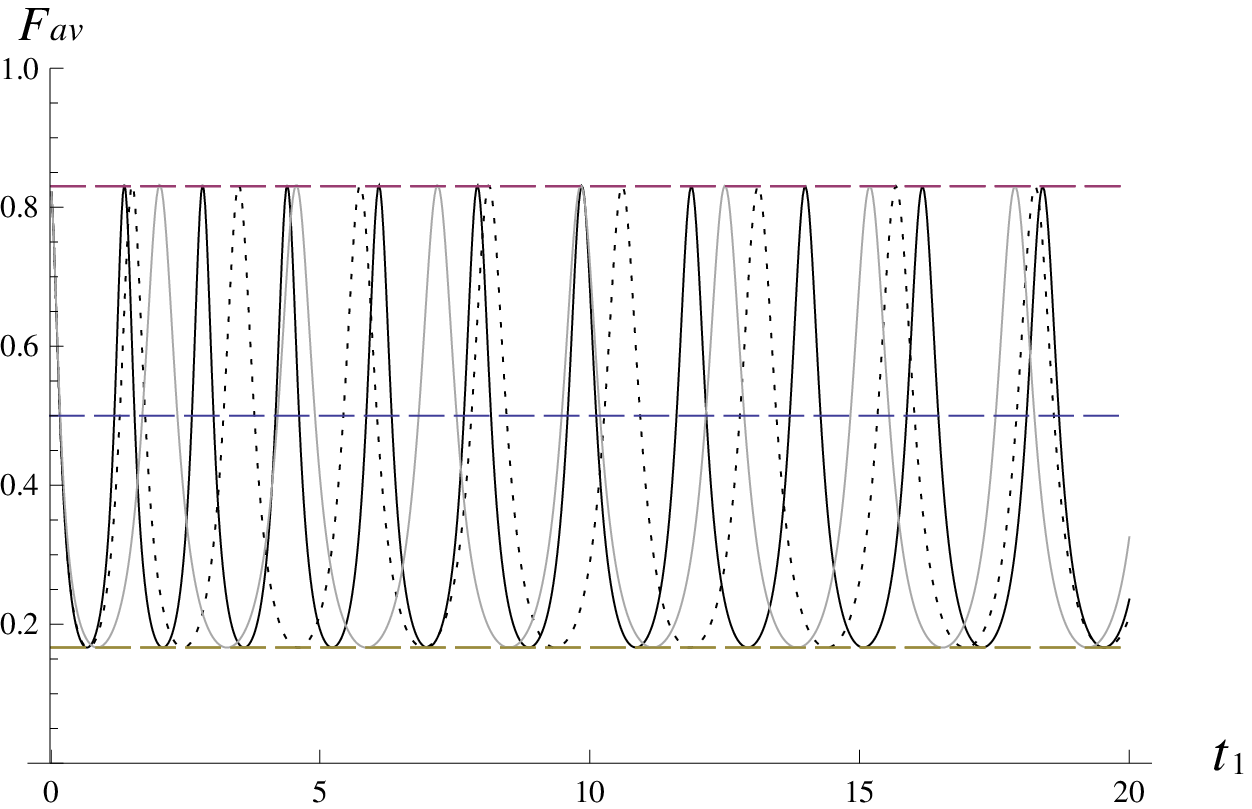}
\includegraphics[width=7.5cm]{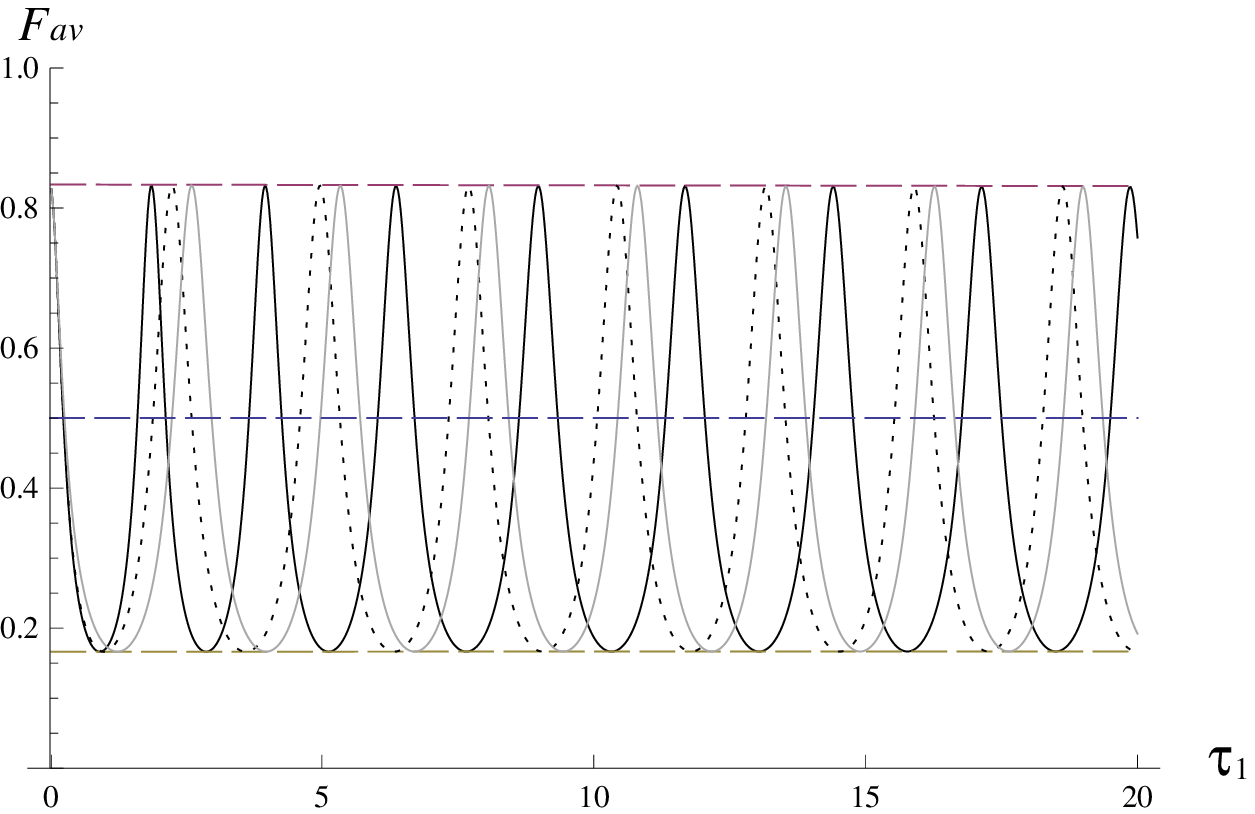}
\caption{Averaged pseudo-fidelity $F_{av}$ in the ultraweak coupling limit in the Minkowski frame (left) and in the quasi-Rindler frame
(right)  at early times for different $a$. Here the parameters are the same as those in Fig. \ref{wcCorr} except that
the proper accelerations are $a=1/4$ (solid), $a=1$ (dotted) and $a=4$ (gray). Dashed lines are $F^+_{av}$ (top), $1/2$ (middle),
and $F^-_{av}$ (bottom), where $F^{\pm}_{av}$ assume the approximated values obtained from $(\ref{maxminFav})$. One can see that
the position of the peaks at early times are different for different $a$. This is because different degrees
of time-dilation of detector $B$ (detector $A$) seen by an observer at rest in Minkowski (quasi-Rindler) frame will shift the
oscillations $\cos \Omega(t_1+\tau(t_1))$ ($\cos \Omega(t(\tau_1)+\tau_1)$) in $(\ref{X2Y2})$ in different ways.}
\label{wcF}
\end{figure}

\begin{figure}
\includegraphics[width=7.5cm]{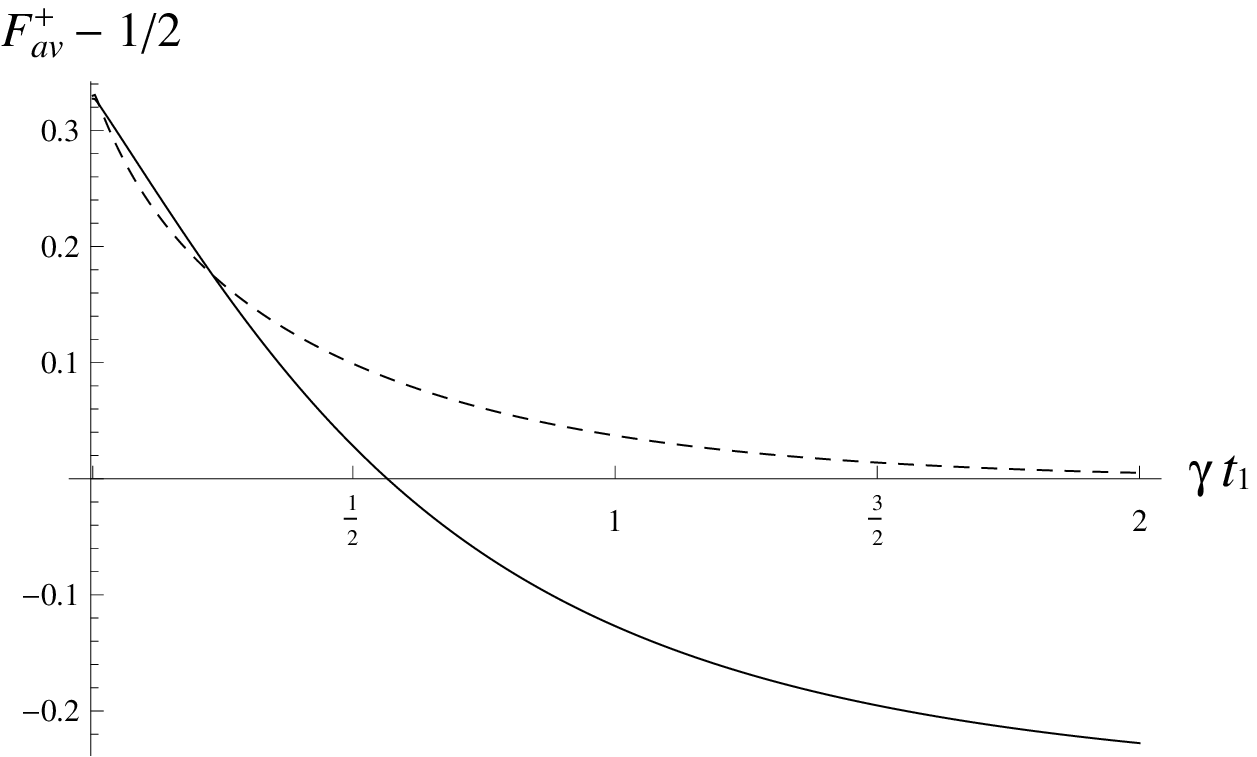}
\includegraphics[width=7.5cm]{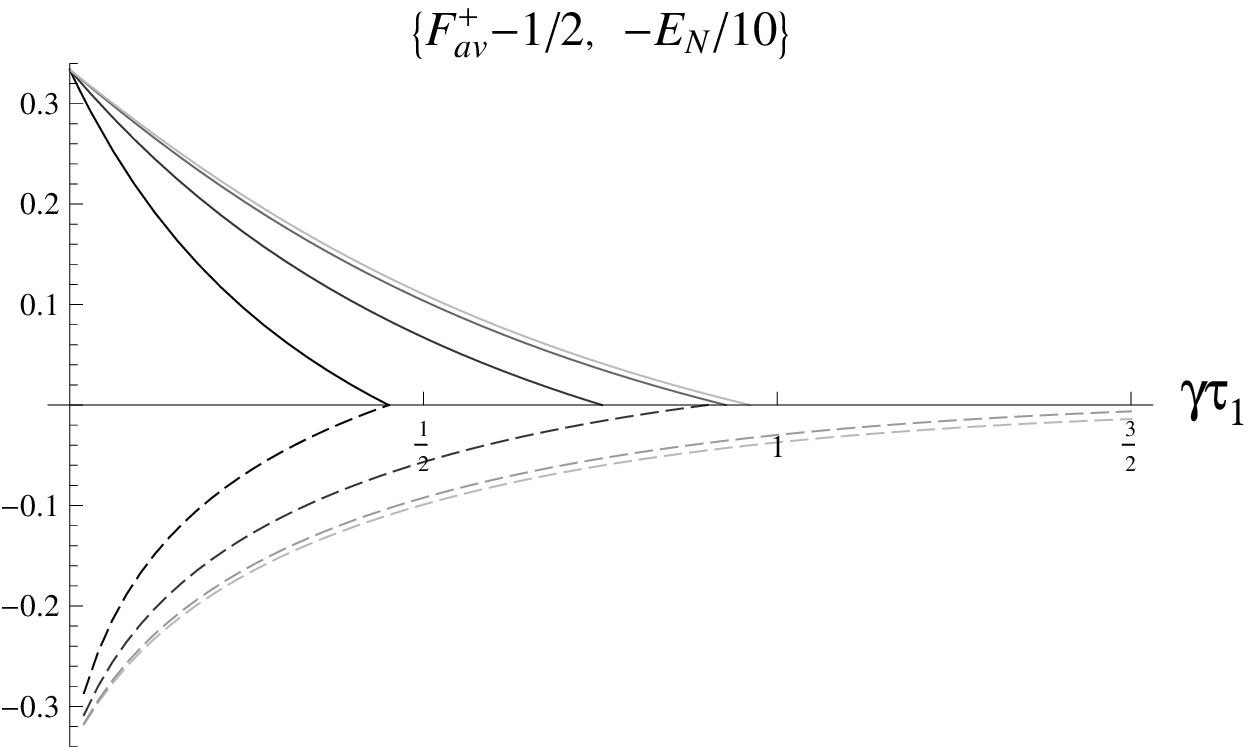}
\caption{(Left) $F^+_{av}(t_1) -1/2$ (solid curves) and $E_{\cal N}(t_1)/10$ (dashed curves) in the Minkowski frame.
The parameters are the same as those in Fig. \ref{wcF}.
The curves for $a=1/4$, $a=1$, and $a=4$ are almost indistinguishable in the left plot since both $F^+_{av}$
and $E_{\cal N}$ are insensitive to $a$ in the Minkowski frame in the ultraweak coupling limit.
One can see that the peak-values of averaged pseudo-fidelity of quantum teleportation, namely $F^+_{av}$, become less than $1/2$ for
$t_1 > t_{1/2} \approx 0.57/\gamma$ when entanglement between $A$ and $B$ are still quite strong right before the joint measurement
in the Minkowski frame [in the sense that the value of the logarithmic negativity $E_{\cal N}(t_1)$ is well above zero].
The disentanglement time is $t_{dE}\approx 3.75/\gamma$ according to Ref. \cite{LCH08}, much later than the moment
$t_{1/2}$ that quantum teleportation loses its advantage.
(Right) $F^+_{av}(t_1) -1/2$ (solid curves) and $-E_{\cal N}(t_1)/10$ (dashed curves) in the quasi-Rindler frame.
There are actually six curves in each set: from right to left they correspond to the cases
with $a=1/4$, $1$, $2$ (all are light gray curves, indistinguishable in this plot), $4$ (gray), $8$ (dark gray), and $16$ (black),
respectively, in quasi-Rindler frame. Other parameters are the same as before. One can see that
$F^+_{av}$ here still fall below the best classical fidelity $1/2$ earlier than the disentanglement time when
$E_{\cal N}$ touches zero. As illustrated here, the larger the proper acceleration $a$, the earlier the value of $F^+_{av}-1/2$
becomes negative in the quasi-Rindler frame. For $a$ large enough, that moment will be quite close to, but no later than the disentanglement
time $\tau_{dE} \approx \pi\Omega/\gamma a$ \cite{LCH08}.}
\label{wcEN}
\end{figure}

Results in the Minkowski frame, where Alice performs the joint measurement at $x^0_{\bf 1}=t_1$ and Rob's proper time
$\tau_1 = \tau(t_1)= a^{-1}\sinh^{-1}at_1$, are shown in Figs. \ref{wcCorr}, \ref{wcF} (left), and \ref{wcEN} (left).
In Figs. \ref{wcCorr} and \ref{wcF} one can see that the averaged pseudo-fidelity $F_{av}$ oscillates in a time-varying
frequency due to the $\cos\Omega(t_1+\tau(t_1))$ term in $(\ref{X2Y2})$. $F_{av}$ is larger as $\left<\right. Q_-^2 \left.\right>$
(and $\left<\right. P_+^2 \left.\right>$) gets smaller, when the two-mode squeezed state of $A$ and $B$ looks closer to the EPR
state so the BK scheme is closer to the idealized case given by Vaidman \cite{BK98, Va94}. In more than half of the period
in an oscillation, however, $F_{av}$ is less than $1/2$ because the squeezing of the $AB$-state has oscillated to the
orthogonal direction, so that $Q_-$ and $P_+$ are uncertain.
One may improve the teleportation by initiating the $AB$-pair as a rotating squeezed state and using a local
oscillator to track its phase angle \cite{QTelepExpts}, or
simply switching to an alternative protocol using $Q_+$ and $P_-$ instead of $Q_-$ and
$P_+$ in Vaidman's scheme whenever $\left<Q_-^2\right> > \left<Q_+^2\right>$ and $\left<P_+^2\right> >\left<P_-^2\right>$,
then $F_{av}$ could be greater than $1/2$ during most of the early times, though the peak values will never exceed $F_{av}^+$.

The positions of the peaks of $F_{av}$ at early times are different in the cases with different proper accelerations $a$ of
detector $B$. This is because in this setup different degrees of time-dilation of Rob seen by an observer at rest in Minkowski
frame will shift the oscillations $\cos \Omega(t_1+\tau(t_1))$ in $(\ref{X2Y2})$ in different ways. As mentioned earlier, since $t$
and $\tau(t)$ depend on the motion of detectors $A$ and $B$, the position of the peaks of $F_{av}$ also depends on
how the detectors move. When $t_1$ gets larger, time dilation of detector $B$ becomes more significant in our setup
and so detector $B$ appears to change extremely slowly in the Minkowski frame.
Thus $F_{av}$ oscillates approximately in frequency $\Omega$ at late times.

The peak values $F^+_{av}$ fall below the best fidelity of classical teleportation $1/2$ at some time much earlier
than the disentanglement time $t_{dE}$ when the logarithmic negativity $E_{\cal N}$ become zero.
One can estimate the moment $t_{1/2}$ when $F_{av}^+$ touches $1/2$ if $a$ is not too small in the ultraweak coupling limit.
For large $t_1$ with $\gamma t_1\sim O(1)$, $\tau(t_1)\ll t_1$, so $e^{-\gamma(t_1+\tau(t_1))}\approx e^{-\gamma t_1}$,
and $B(t) \approx 2+C_2 + C_1$. From $(\ref{maxminFav})$, one has
\begin{equation}
  F_{av}^+(t_1) \approx {2{\cal A}(t_1)\over(2+C_2+C_1){\cal A}(t_1)-(S_2+e^{-\gamma t_1}S_1)^2}
  \label{FavpMwc}
\end{equation}
where ${\cal A}(t_1)$ has been given in $(\ref{Aoft})$, which is independent of $a$ in this limit.
So one can see that, while the moments when $F_{av}(t_1)$ reaches a local extremum depend on
the proper acceleration $a$ of detector $B$,
$F^+_{av}(t_1)$ in the ultraweak coupling limit is insensitive to $a$ 
in the Minkowski frame, just like the degree of entanglement $\Sigma$ or $E_{\cal N}$ in this case.
From $(\ref{FavpMwc})$, one obtains $t_{1/2}$ by solving the equation
\begin{equation}
   (C_1-1)(C_2-3)e^{-2\gamma t_{1/2}}-2S_1 S_2 e^{-\gamma t_{1/2}} - (C_1 -1)(C_2+1) = 0,
\end{equation}
such that $F_{av}^+(t_{1/2}) \approx 1/2$.
Unlike the disentanglement time in this case \cite{LCH08}, $t_{dE}\approx (2\gamma)^{-1}\ln (\pi\Omega/\gamma\Lambda_1)$,
which is almost independent of the initial state of the entangled detectors,  the moment $t_{1/2}$ when $F_{av}^+
\approx 1/2$ strongly depends on the squeezed parameter $r_1$ of the initial state of detectors $A$ and $B$, as well
as the squeezed parameter $r_2$ introduced by the joint measurement on $A$ and $C$, though $t_{1/2}$ is still insensitive
to $a$.

\subsubsection{Quasi-Rindler frame}

By a quasi-Rindler frame we refer to the coordinate system in which each time-slice almost overlaps a Rindler time-slice in
the R-wedge but the part in the L-wedge has been bent to the region with positive $t$ to make the whole time-slice located
after the initial time-slice for the Minkowski observer, as illustrated in Fig. \ref{AR}.

Results in the quasi-Rindler frame in the ultraweak coupling limit, where Alice performs the joint measurement at $x_{\bf 1}^0 = \tau_1$
such that $t_1 = t(\tau_1)=b^{-1}\tanh a\tau_1$, are shown in Figs. \ref{wcF} (right) and \ref{wcEN} (right). In Fig. \ref{wcF} (right)
there are similar oscillations to those in Fig. \ref{wcF} (left) because of the same $\cos \Omega(t_1(\tau_1)+\tau_1)$ term in
$(\ref{X2Y2})$, but here the shift of the peaks at early times is due to the time-dilation of detector $A$ seen by the Rindler
observer. When $\tau_1$ gets larger, the frequency of the oscillation also approaches $\Omega$, since detector $A$ looks
frozen in the quasi-Rindler frame.

In contrast to the case in the Minkowski frame, $F^+_{av}(\tau_1)$ in the quasi-Rindler frame is sensitive to the proper acceleration $a$.
Indeed, if $a$ and $b$ are not extremely small, for large $\tau_1$, one has $t_1 = t(\tau_1) \approx 1/b$,
\begin{equation}
  F_{av}^+(\tau_1) \approx {2\left[C_2+1+e^{-2\gamma/b}(C_1-1)\right]\over
  \left[C_2+1+e^{-2\gamma/b}(C_1-1)\right]{\cal B}(\tau_1)-(S_2+e^{-\gamma(\tau_1+1/b)}S_1)^2},
  \label{FavRindwc}
\end{equation}
where ${\cal B}(\tau_1)$ given in $(\ref{BofT})$ depends on $a$. Again, the moment $\tau_{1/2}$ when $F_{av}^+(\tau_{1/2})=1/2$
depends on $r_1$ in the initial state of the detectors $A$ and $B$ as well as $r_2$ from the joint measurement on $A$ and $C$.

In Fig. \ref{wcEN} (right) one can see that the larger proper acceleration $a$, the earlier the value of $F^+_{av}-1/2$ touches
zero, while the value of $\tau_{1/2}$ is always less than $\tau_{dE}$ ($\approx \pi\Omega/\gamma a$ for large $a$ \cite{LCH08}).
Indeed, the conclusion in Appendix A is valid for the Rindler observer as well as the Minkowski observer:
it implies that the peak values of the averaged pseudo-fidelity of quantum teleportation $F^+_{av}$ must have been less than
the best averaged fidelity $1/2$ of classical teleportation at the disentanglement time $\tau_{dE}$ when $E_{\cal N}$ touches zero, so
$\tau_{1/2} \le \tau_{dE}$. 
For large $a$, this can be easily seen by inserting $\tau_1=\tau_{dE}\approx\pi\Omega/\gamma a$ into $(\ref{FavRindwc})$,
which implies that $F_{av}^+ (\tau_{dE}) \ge 1/2$ if
\begin{eqnarray}
  0 &\ge& (C_1+1)(C_2+1)e^{-2\pi\Omega/a} - 2 e^{-\gamma/b}S_1 S_2 e^{-\pi\Omega/a} + e^{-2\gamma/b}(C_1-1)(C_2-1) \nonumber\\
  &=& (C_1+1)(C_2+1)\left( e^{-\pi\Omega/a} - e^{-\gamma/b} \tanh r_1 \tanh r_2\right)^2.
\end{eqnarray}
But the right hand side of the above equation is positive definite. Thus for all parameters $r_1$, $r_2$, and $\Omega$,
$F_{av}^+(\tau_{dE})$ is always less than $1/2$, unless the parameters happen to satisfy the equality
\begin{equation}
  {\pi\Omega\over a} = {\gamma\over b} - \ln (\tanh r_1 \tanh r_2)
\end{equation}
such that $\tau_{1/2}\approx \tau_{dE}$ in this particular case.

\subsection{Beyond ultraweak coupling limit}

Beyond the ultraweak coupling limit, both $F_{av}$ and $E_{\cal N}$ are strongly affected by the environment.
The calculation can be more complicated if the mutual influences between detectors $A$ and $B$ are strong.
For simplicity, we consider the cases with $b>2a$ with $a$ not extremely large so that only the first and second
order corrections from the mutual influences are needed while they are still small compared with the zeroth order \cite{LCH08}.

\begin{figure}
\includegraphics[width=7.5cm]{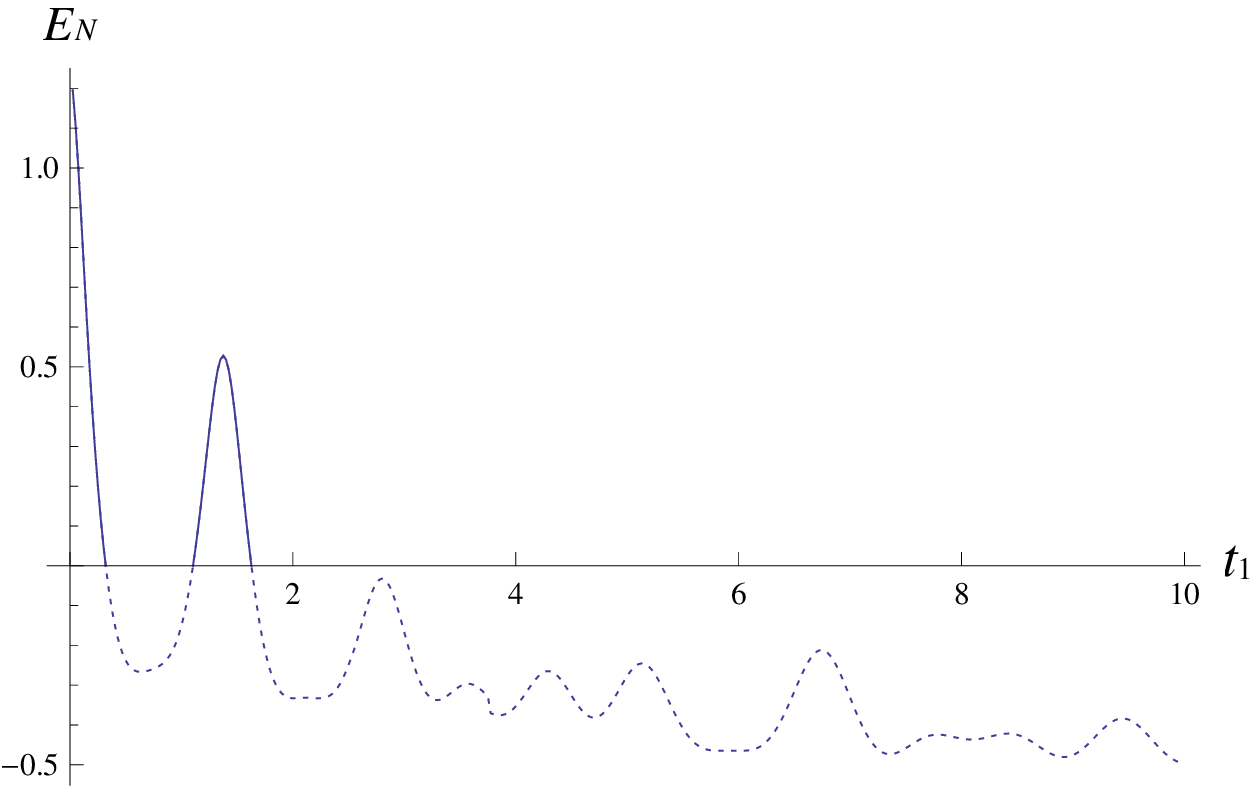}
\includegraphics[width=7.5cm]{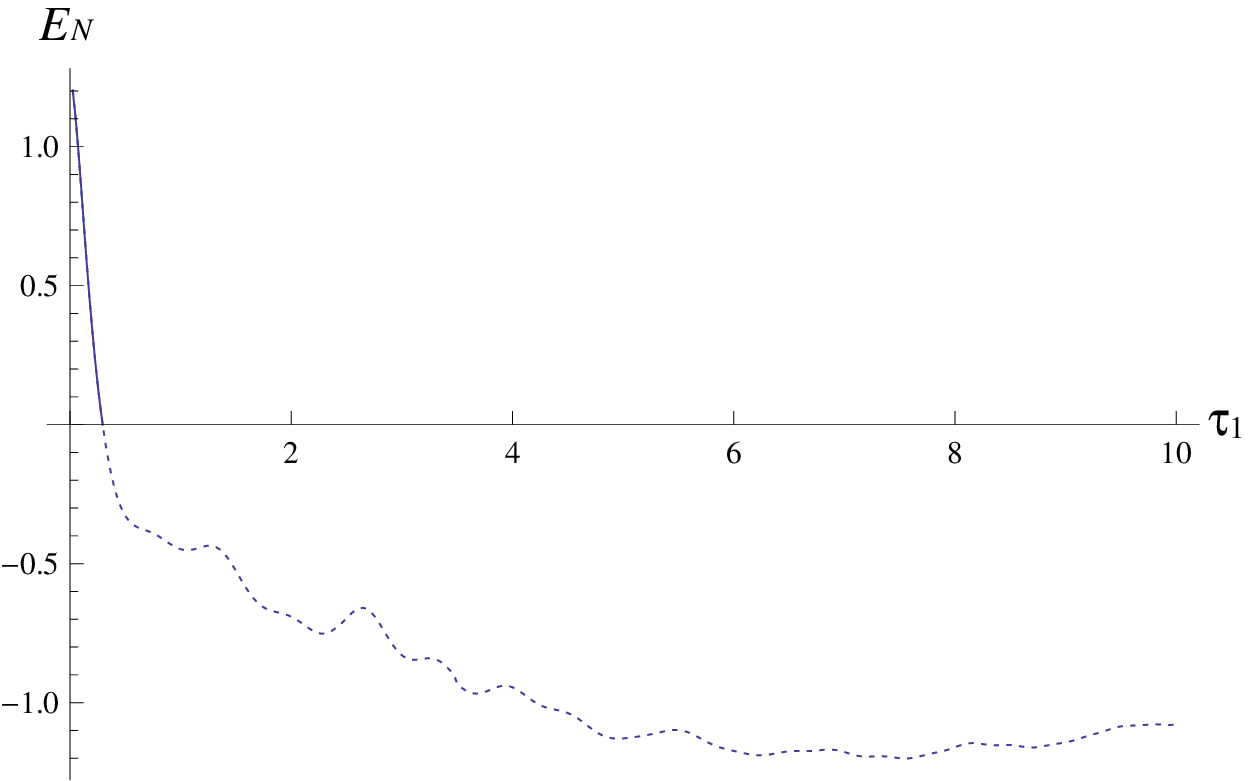}\\
\includegraphics[width=7.5cm]{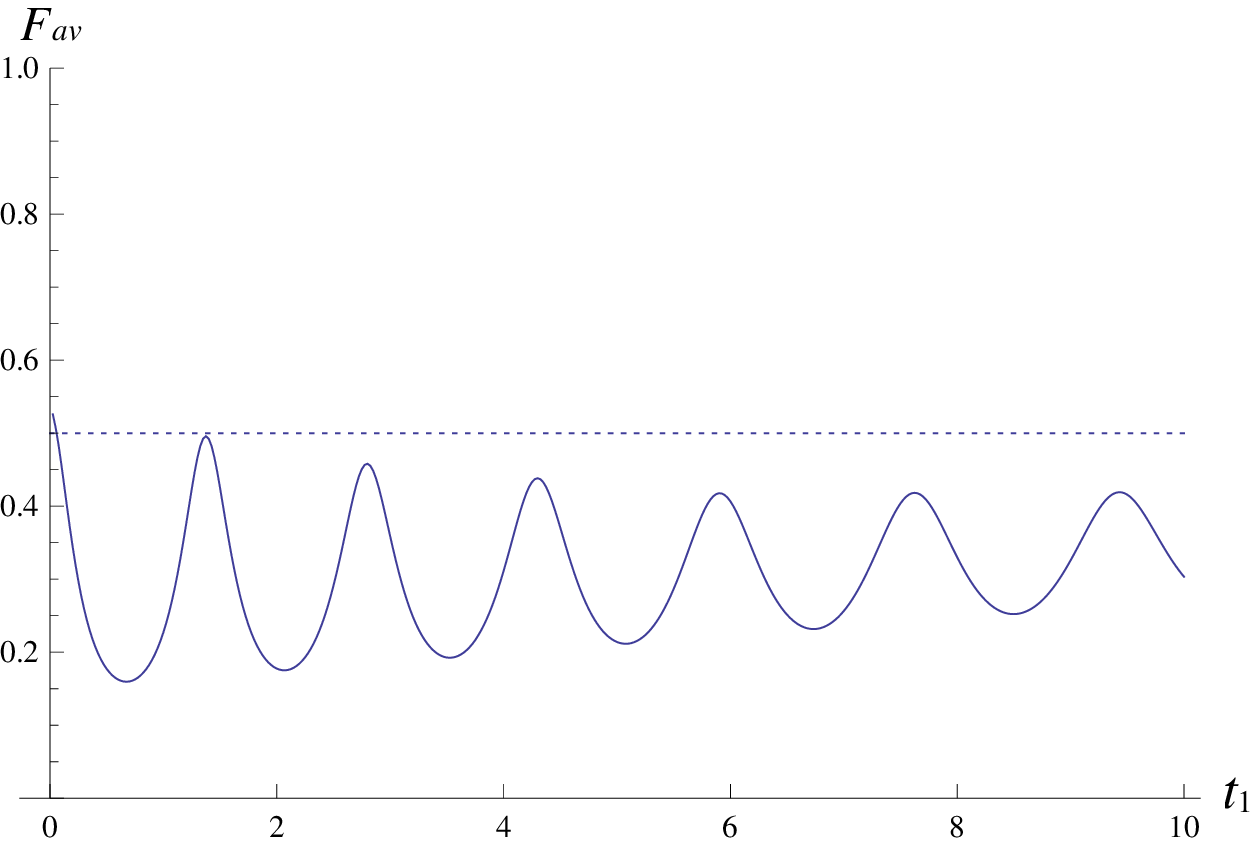}
\includegraphics[width=7.5cm]{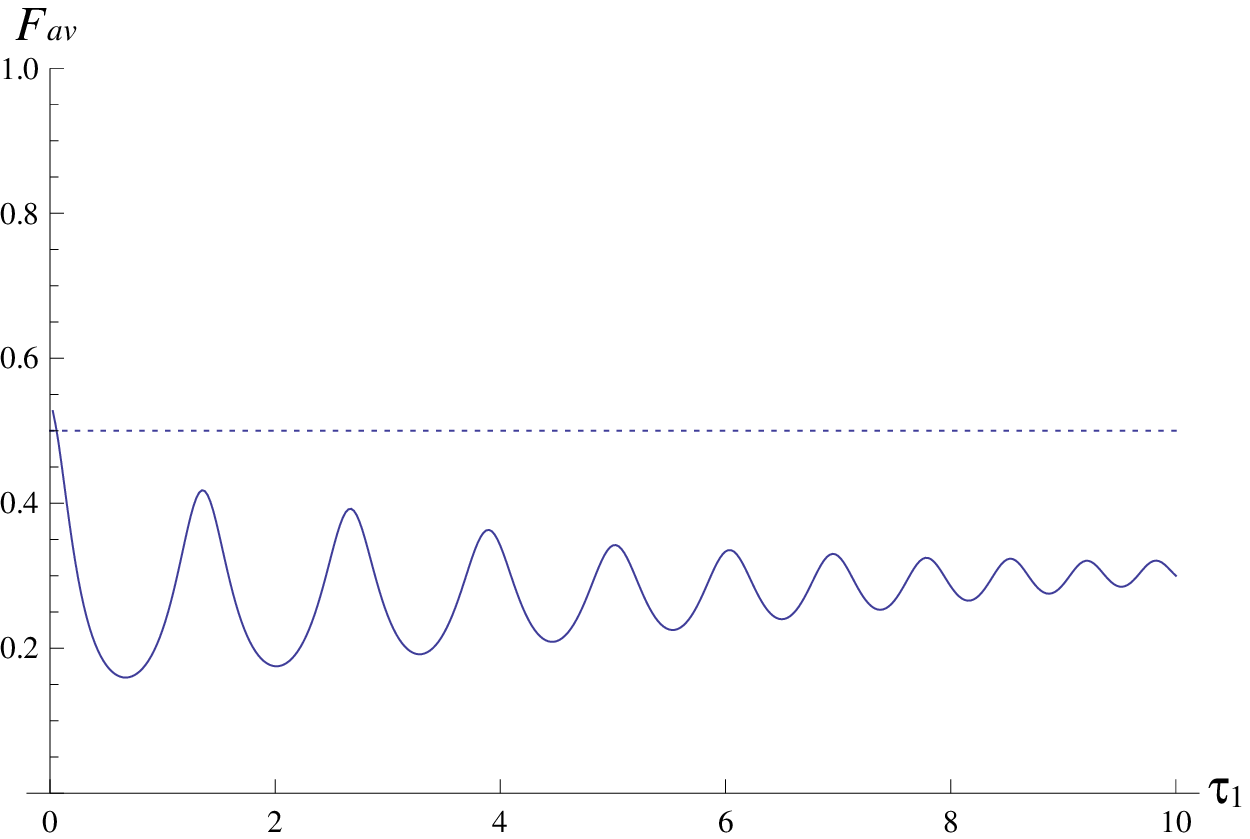}
\caption{Beyond the ultraweak coupling limit, quantum entanglement disappears quickly both in the Minkowski frame (upper left)
and in the quasi-Rindler frame (upper right) as witnessed by $E_{\cal N}$ (solid curves) becoming zero quickly.
The averaged pseudo-fidelity $F_{av}$ drops below $1/2$ even quicker, both in the Minkowski frame (lower left)
and in the quasi-Rindler frame (lower right).
Here $\gamma=0.1$, $\Omega=2.3$, $(a, b)=(0.2,0.401)$, $r_1=1.1$, $r_2=1.2$, and $\Lambda_0=\Lambda_1=20$. }
\label{StrongCoup}
\end{figure}

In Fig. \ref{StrongCoup} one can see that quantum entanglement disappears quickly both in the Minkowski frame and the quasi-Rindler frame
due to strong interplays with the environment.
The averaged pseudo-fidelity $F_{av}$ drops below $1/2$ even quicker, and the peak values of $F_{av}$ never exceed $1/2$
again once they dropped below this level in these examples.

\section{Physical fidelity in a more realistic setup}
\label{physreal}

\begin{figure}
\includegraphics[width=6cm]{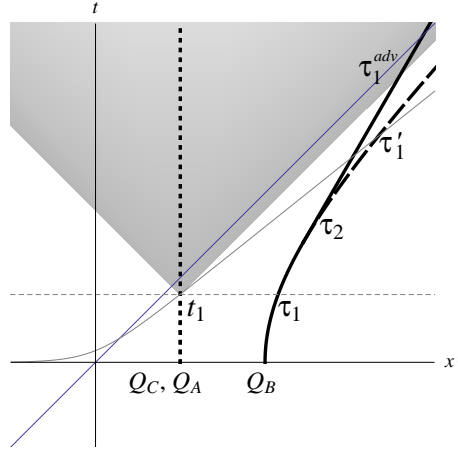}
\caption{Setup for quantum teleportation from Alice (thick dotted line) at rest to Rob (thick solid curve)
accelerated constantly from $0$ to $\tau_2$ in his proper time then turning to inertial motion.
The gray solid curve represents the $\tau_1'$-slice in the quasi-Rindler frame, and the gray dotted horizontal line
represents the $t_1$-slices in the Minkowski frame.
The shaded region represents the future lightcone of the joint measurement event on $A$ and $C$ by Alice, and
the hypersurface $t=x^1$ is the event horizon of Rob for $\tau_2\to\infty$.}
\label{AR}
\end{figure}

Suppose Rob stops accelerating at his proper time $\tau_2$ when $t_2 = a^{-1}\sinh a\tau_2$ in Minkowski time,
after this moment Rob moves with constant velocity along the worldline $( (\tau-\tau_2)\cosh a \tau_2+a^{-1}\sinh a\tau_2,
(\tau-\tau_2)\sinh a \tau_2 + a^{-1}\cosh a\tau_2, 0,0)$ in Minkowski coordinate for $\tau^{}_B=\tau>\tau_2$, while Alice stays 
at $(t, 1/b, 0,0)$ and performs the joint measurement on A and C at $t_1$ (see Fig. \ref{AR}). In this setup the classical information
about the outcome at the very moment of the measurement can always reach Rob if the signal is traveling at the speed of light.
Assume Alice sends out the information right after $t_1$ when the joint measurement on $A$ and $C$ is done, then Rob will
receive the message at his proper time
\begin{equation}
  \tau^{adv}_1 =\left\{ \begin{array}{lll}
   -a^{-1}\ln a\left( b^{-1} - t_1\right)  & \,\,\,{\rm if}\,\,\,  t_1 < b^{-1}-a^{-1}e^{-a\tau_2}, \\
   \left(t_1 - b^{-1}\right)e^{a\tau_2} + a^{-1} + \tau_2 & \,\,\,{\rm otherwise.}    \end{array}  \right.
\label{tau1adv}
\end{equation}

Suppose Rob performs the local operation on $B$ at some moment $\tau^{}_P > \tau^{adv}_1$ according to the received information.
Then the averaged physical fidelity should be given by
\begin{equation}
  F_{av} = \int d^2\beta P(\beta)\times{}^{}_B\left<\alpha\right| \hat{\rho}^{}_{out}(\tau^{}_P)\left|\alpha\right>^{}_B,
\end{equation}
where $\rho^{}_{out}(\tau^{}_P)$ has the same form as Eq.($\ref{rhoOut}$) but $\tilde{\rho}^{}_B$ defined at $t_1$ there is
replaced by $\tilde{\rho}^{}_B(\tau^{}_P)$ which started with the initial state $\tilde{\rho}^{}_B(\tau_1)$ with
$\tau_1 = a^{-1}\sinh^{-1}a t_1$ and evolves from $\tau_1$ to $\tau^{}_P$ according to the Schr\"odinger equation.
During the continuous evolution mutual influences from detector $A$ will start to affect $B$ at $\tau_1^{adv}$.
This makes the calculation more complicated. Fortunately, if the classical signal travels at lightspeed and Rob performs
the local operation right after he receives the signal, namely, $\tau^{}_P = \tau^{adv}_1 + \epsilon$ with $\epsilon\to 0+$,
and if mutual influences are weak enough and negligible (this is easy to satisfy in the weak coupling and large distance limit,
see \cite{OLMH11}), the calculation can be greatly simplified.

\subsection{Reduced state of a detector with its entangled partner being measured}

In Ref. \cite{Lin11a} one of us has shown that a quantum state of a Raine-Sciama-Grove detector-field system in (1+1)D Minkowski space started with the same initial state defined on the same fiducial time-slice, then collapsed by a {\it spatially local}
measurement on the detector at some moment, will evolve to the same quantum state on the same final time-slice (up to a coordinate
transformation), independent of which frame is used by the observer or which time-slice the wave function collapsed on
between the initial and the final time-slices. This implies that the reduced state of the detector $B$ at the final time is coordinate-independent. 
For the Unruh-DeWitt detector theory in (3+1)D Minkowski space considered here, the argument is similar.

Right after the local measurement on detectors $A$ and $C$ at $t_1$ (for a simpler case with the local measurement
only on detector $A$, see Ref. \cite{Lin11b}), the quantum state at $t_1$ 
collapses to $\tilde{\rho}^{}_{AC} \otimes \tilde{\rho}^{}_{B\Phi_{\bf x}}$ on $t_1$-slice in the Minkowski frame or
$\tau_1'$-slice in quasi-Rindler frame (see Fig. \ref{AR}), or whatever time-slice depending on the observer's frame.
Similar to $(\ref{rhoB})$, here the post-measurement state
$\tilde{\rho}^{}_{B\Phi_{\bf x}}$ of detector $B$ and the field $\Phi_{\bf x}$ is obtained by
\begin{eqnarray}
  \tilde{\rho}^{}_{B\Phi_{\bf x}}(K^{\bar{\sigma}},\Delta^{\bar{\sigma}}) &=& N \int {dK^C d\Delta^C \over 2\pi\hbar}
  {dK^A d\Delta^A \over 2\pi\hbar}\tilde{\rho}_{AC}^*(K^A, K^C, \Delta^A,\Delta^C)\rho({\bf K}, K^{\bf x}, {\bf \Delta},\Delta^{\bf x}; t_1)
\label{rhoBPhi0}
\end{eqnarray}
where $\rho$ is the quantum state of the combined system evolved from $t_0\equiv 0$ to $t_1$ and ${\bar{\sigma}}
=\{ B\}\cup \{ {\bf x} \}$. Since $\tilde{\rho}^{}_{AC}$ is Gaussian, a straightforward
calculation shows that the post-measurement state of detector $B$ and the field has the form:
\begin{eqnarray}
  \rho^{}_{B\Phi_{\bf x}}(K^{\bar\sigma},\Delta^{\bar{\sigma}})
  &=& \exp \left[ {i\over \hbar} \left( {\cal J}^{(0)}_{\bar{\zeta}} K^{\bar{\zeta}} -{\cal M}^{(0)}_{\bar{\zeta}}\Delta^{\bar{\zeta}}\right)
  -{1\over 2\hbar^2} \left( K^{\bar{\zeta}} {\cal Q}_{{\bar{\zeta}}{\bar{\xi}}} K^{\bar{\xi}} +
  \Delta^{\bar{\zeta}} {\cal P}_{{\bar{\zeta}}{\bar{\xi}}} \Delta^{\bar{\xi}}-
  2 K^{\bar{\zeta}} {\cal R}_{{\bar{\zeta}}{\bar{\xi}}} \Delta^{\bar{\xi}} \right) \right.\nonumber\\
  & & \hspace{.5cm} \left.  +{1\over 2\hbar^2}\sum_{n=1}^4 {1\over {\cal W}^{(n)}}
  \left( K^{\bar{\zeta}}{\cal J}^{(n)}_{\bar{\zeta}}- \Delta^{\bar{\zeta}}{\cal M}^{(n)}_{\bar{\zeta}} \right)
  \left( {\cal J}^{(n)}_{\bar{\xi}} K^{\bar{\xi}}- {\cal M}^{(n)}_{\bar{\xi}}\Delta^{\bar{\xi}}\right) \right].
\label{rhoBPhi}
\end{eqnarray}
Here we use the Einstein-DeWitt notation for $\bar{\zeta}, \bar{\xi} =\{ B\}\cup \{ {\bf x} \} $, which run
over the degrees of freedom of detector $B$ and the field defined at ${\bf x}$ on the whole time-slice,
$n$ runs from $1$ to $4$ corresponding to the four dimensional Gaussian integrals in $(\ref{rhoBPhi0})$,
${\cal W}^{(n)}$ depends only on the two-point correlators of detectors $A$ and $C$ at the moment of measurement,
while ${\cal J}^{(i)}_{\bar{\zeta}}(\hat{\Phi}_{\bar{\zeta}})$ and ${\cal M}^{(i)}_{\bar{\zeta}}(\hat{\Pi}_{\bar{\zeta}})$ are linear
combinations of the terms with a cross correlator between detector $A$ or $C$ and the operators
$\hat{\Phi}_{\bar{\zeta}}$ or $\hat{\Pi}_{\bar{\zeta}}$ ($\hat{\Phi}_{B}\equiv \hat{Q}_B$ and $\hat{\Pi}_{B}\equiv \hat{P}_B$),
respectively, multiplied by a few correlators of $A$ and/or $C$,
all of which are the correlators of the operators evolved from $t_0$ to $t_1$ with respect to the initial state given at $t_0$.          
This implies that the two-point correlators right after the wave functional collapse become
\begin{eqnarray}
  \left<\right. \hat{\Phi}^{[1]}_{\bar{\zeta}},\hat{\Phi}^{[1]}_{\bar{\xi}}\left.\right>_1
    &=& \left<\right.\hat{\Phi}^{[10]}_{\bar{\zeta}}, \hat{\Phi}^{[10]}_{\bar{\xi}}\left.\right>_0 -
    \sum_{n=1}^4  {{\cal J}^{(n)}_{\bar{\zeta}}(\hat{\Phi}^{[10]}_{\bar{\zeta}})
    {\cal J}^{(n)}_{\bar{\xi}}(\hat{\Phi}^{[10]}_{\bar{\xi}}) \over {\cal W}^{(n)}}, \label{Q2PM} \\
  \left<\right. \hat{\Pi}^{[1]}_{\bar{\zeta}},\hat{\Pi}^{[1]}_{\bar{\xi}}\left.\right>_1
    &=& \left<\right.\hat{\Pi}^{[10]}_{\bar{\zeta}}, \hat{\Pi}^{[10]}_{\bar{\xi}}\left.\right>_0 -
    \sum_{n=1}^4  {{\cal M}^{(n)}_{\bar{\zeta}}(\hat{\Pi}^{[10]}_{\bar{\zeta}})
    {\cal M}^{(n)}_{\bar{\xi}}(\hat{\Pi}^{[10]}_{\bar{\xi}}) \over {\cal W}^{(n)}}, \label{P2PM}\\
  \left<\right. \hat{\Phi}^{[1]}_{\bar{\zeta}},\hat{\Pi}^{[1]}_{\bar{\xi}}\left.\right>_1
    &=& \left<\right.\hat{\Phi}^{[10]}_{\bar{\zeta}}, \hat{\Pi}^{[10]}_{\bar{\xi}}\left.\right>_0 -
    \sum_{n=1}^4  {{\cal J}^{(n)}_{\bar{\zeta}}(\hat{\Phi}^{[10]}_{\bar{\zeta}})
    {\cal M}^{(n)}_{\bar{\xi}}(\hat{\Pi}^{[10]}_{\bar{\xi}}) \over {\cal W}^{(n)}}. \label{PQPM}
\end{eqnarray}
For example,
$\left<\right.(\hat{Q}_B^{[1]})^2\left.\right>_1= {\cal Q}_{BB}(t_1)  -
\sum_{n=1}^4 [{\cal J}^{(n)}_{B}(\hat{Q}_B^{[10]}) {\cal J}^{(n)}_{B}(\hat{Q}_B^{[10]})/{\cal W}^{(n)}]$
where ${\cal Q}_{BB}(t_1) =\left<\right.(\hat{Q}_B^{[10]})^2\left.\right>_0$.
Here $\hat{\cal O}_B^{[1]}$ refers to the operator $\hat{\cal O}_B$ defined at $t_1$ and
$\hat{\cal O}_B^{[10]}$ refers to the operator $\hat{\cal O}_B(t_1-t_0)$ in the Heisenberg picture.

At some moment $t_M$ in the Minkowski frame before the detector $B$ enters the future lightcone of the measurement event on $A$, namely,
when $\tau^{}_B=\tau(t_M) < \tau^{adv}_1$, the two-point correlators of the detector $B$ is either in the original, uncollapsed form,
e.g. $\left<\right. \hat{Q}_B^2(t_M-t_0) \left.\right>_0$, if the wave functional collapse does not happen yet in some observers'
frames, or in the collapsed form evolved from the post-measurement state, e.g.,
\begin{eqnarray}
  \left<\right. \hat{Q}_B^2(t_M) \left.\right> &=&
  \left< \left[ \sum_{{\bf d}} \left(\phi_B^{{\bf d}[M1]}\hat{Q}_{\bf d}^{[1]}+
  f_B^{{\bf d}[M1]}\hat{P}_{\bf d}^{[1]}\right)
  +\int dx \left(\phi_B^{x[M1]}\hat{\Phi}_x^{[1]}+f_B^{x[M1]}\hat{\Pi}_x^{[1]}\right) \right]^2 \right>_1 \nonumber\\
 & & = \left<\right. (\tilde{\Upsilon}_B^{[M0]})^2 \left. \right>_0 -\sum_{n=1}^4
  {{\cal I}^{(n)} [\tilde{\Upsilon}_B^{[M0]}, \tilde{\Upsilon}_B^{[M0]} ]\over {\cal W}^{(n)}} ,
\label{QB2clpsed}
\end{eqnarray}
in other observers' frames. Here we have used the Huygens' principles $(\ref{id1})$ and $(\ref{id2})$, and defined
\begin{eqnarray}
  \tilde{\Upsilon}^{[M0]}_B &\equiv& \hat{\Phi}_\zeta^{[0]}\left[\phi_B^{\zeta[M0]}-\phi_B^{A[M1]}\phi_A^{\zeta[10]}-
  f_B^{A[M1]}\pi_A^{\zeta[10]}\right] + \hat{\Pi}_\zeta^{[0]}\left[f_B^{\zeta[M0]}-\phi_B^{A[M1]}f_A^{\zeta[10]}-
  f_B^{A[M1]}p_A^{\zeta[10]}\right] 
\label{Updef}
\end{eqnarray}
with $\hat{\Phi}_{A,C}\equiv \hat{Q}_{A,C}$ and $\hat{\Pi}_{A,C}\equiv \hat{P}_{A,C}$,
while ${\cal I}^{(n)}$ is derived from those ${\cal J}^{(n)}_{\bar{\zeta}}$ and ${\cal J}^{(n)}_{\bar{\xi}}$  pairs in
$(\ref{Q2PM})$-$(\ref{PQPM})$.
Note that before the detector $B$ enters the lightcone, $\phi_B^{A[M1]} = f_B^{A[M1]}=0$, such that
$\tilde{\Upsilon}^{[M0]}_B$ reduces to $\hat{Q}^{[M0]}_B$.
So at the moment $t_M$ the correlators of detector $B$ do not depend on the data on $t_1$-slice except those right
at the local measurement event on detector $A$ and $C$. This means that once we discover the reduced state of detector $B$ 
has been collapsed, the form of the reduced state of $B$ will be independent of the moment when the collapse occurs
in the history of detector $B$ (e.g. $\tau^{}_B=\tau_1$ or $\tau'_1$ in Fig. \ref{AR} if $\tau_2 > \tau'_1$ there),
namely, the moment where the worldline of detector $B$ intersects the time-slice that the wave functional collapsed on.

No matter in which frame the system is observed, the correlators in the reduced state of detector $B$ must have become
the collapsed ones like $(\ref{QB2clpsed})$ exactly when detector $B$ is entering the future lightcone of
the measurement event by Alice, namely, $\tau^{}_B = \tau^{adv}_1$, 
after which the reduced states of detector $B$ observed in different frames become consistent.
Also after this moment the retarded mutual influences will reach $B$ such that $\phi_B^{A[M1]}$ and $f_B^{A[M1]}$ would become nonzero and
get involved in the correlators of $B$. In fact, some information of measurement has entered the correlators of $B$ via the
correlators of $A$ and $C$ at $t_1$ at the position of Alice in ${\cal J}^{(n)}$, ${\cal M}^{(n)}$ and ${\cal W}^{(n)}$ much earlier.
Nevertheless, that information cannot be recognized by Rob before he has causal contact with Alice. A short discussion
on this point is given in Appendix B.

Thus we are allowed to choose $t_M$ in $(\ref{QB2clpsed})$ so that $\tau^{}_M\equiv \tau^{}_B(t_M) = \tau^{adv}_1-\epsilon$
and collapse or project the wave functional right before $\tau^{}_M$, namely, collapse on a time-slice almost overlapping
the future lightcone of the measurement event by Alice.
It is guaranteed that there exists some coordinate system having such a time slice which intersects the worldline of Alice
at $\tau^{}_A = t_1$ and the worldline of Rob at $\tau^{}_B = \tau^{adv}_1-\epsilon$.

If we further assume that mutual influences are negligible and Rob performs the local operation right after the classical
information from Alice is received, namely, $\tau^{}_P = \tau^{adv}_1+\epsilon$ with $\epsilon\to 0+$, then
the continuous evolution of the reduced state of detector $B$ from $\tau^{}_M$ to $\tau^{}_P$
is negligible. In this case we can still calculate the averaged fidelity of quantum teleportation using $(\ref{Favformula})$
with $(\ref{NormB})$ and $(\ref{tildeV})$ by inserting $\tau^{adv}_1$ into $\tau_1$ there, then working out the correlators
${\cal S}^{[1]}_{BB}$, ${\cal S}^{[1]}_{AB}$, and ${\cal S}^{[1]}_{BA}$ (${\cal S}= {\cal Q}, {\cal P}, {\cal R}$) accordingly.

\subsection{Correlators of non-uniformly accelerated detector}

To guarantee the classical information from Alice can always reach Rob, we have assumed Rob
stops accelerating at the moment $\tau_2$. This means the acceleration of detector $B$ is not uniform.

The dynamics of the correlators of non-uniformly accelerated detectors have been studied in Ref. \cite{OLMH11}.
In the weak coupling limit the behavior of such a detector is similar to a harmonic oscillator in contact with a heat bath with  a time-varying ``temperature" corresponding to the proper acceleration of the detector.
From Ref. \cite{OLMH11}, the dynamics of entanglement will be dominated by the zeroth order results of
the a-parts of the self  and cross correlators and the v-parts of the self correlators of detectors $A$ and $B$.
The deviation of the v-parts of the self correlators of detector $B$ from those of a uniformly accelerated detector
(an inertial detector in \cite{OLMH11}) and higher-order corrections from mutual influences are negligible in the weak coupling limit with
large initial entanglement and large spatial separation between the detectors.

For larger initial accelerations of detector $B$, the changes of the v-part of its self correlators during and after the transition of
the proper acceleration of detector $B$ from $a$ to $0$ are more significant.
Suppose the changing rate of the proper acceleration of detector $B$ from a finite $a$ to $0$ is fast enough so that we can
approximate the proper acceleration of detector $B$ as a step function of time, but not too fast to produce significant
non-adiabatic oscillation on top of the smooth variation. According to the results in \cite{OLMH11} and \cite{LinDICE10},
for $\tau_2$ sufficiently large, the correlators of detector $B$ behave roughly like
\begin{eqnarray}
  \left<\right.Q_B^2(\tau)\left.\right>_{\rm v} &\approx&
    \left.\left<\right.Q_B^2(\tau)\right|_{a_\mu a^\mu = a^2}\left.\right>_{\rm v} +
    \theta(\tau-\tau_2)\times\nonumber\\ & & \left[
    \left( \left.\left<\right.Q_B^2(\infty)\right|_0 \left.\right>_{\rm v}
    -\left.\left<\right.Q_B^2(\infty)\right|_{a^2}\left.\right>_{\rm v}\right)
    \left(1-e^{-2\gamma (\tau-\tau_2)}\right)
    -{\gamma\hbar a^2 e^{-2\gamma (\tau-\tau_2)}\over 6\pi m_0(\gamma^2+\Omega^2)^2}\right],
    \label{QB2NUAD} \\
  \left<\right.P_B^2(\tau)\left.\right>_{\rm v} &\approx&
    \left.\left<\right.P_B^2(\tau)\right|_{a^2}\left.\right>_{\rm v} +
    \theta(\tau-\tau_2)\left[
    \left( \left.\left<\right.P_B^2(\infty)\right|_0\left.\right>_{\rm v}
    -\left.\left<\right.P_B^2(\infty)\right|_{a^2}\left.\right>_{\rm v}\right)
    \left(1-e^{-2\gamma (\tau-\tau_2)}\right)\right],
    \label{PB2NUAD}
\end{eqnarray}
where $\left<\right.Q_B^2(\infty)\left.\right>_{\rm v}$ and $\left<\right.P_B^2(\infty)\left.\right>_{\rm v}$
are the correlators in steady state at late times.
These approximated bahaviors have been verified by numerical calculations (see Figs. 3(right) and 4(right) in \cite{LinDICE10}).
Note that the last term of $(\ref{QB2NUAD})$ is actually $O(\gamma)$ and negligible in the ultraweak coupling limit.
Also $\left<Q_A^2\right>_{\rm v}$ and $\left<P_A^2\right>_{\rm v}$ behave as  the approximated form in $(\ref{tildeVwc})$,
and other $\left<\right. \cdots \left.\right>_{\rm v}$ are $O(\gamma)$ and negligible in the this limit.
Below we apply these approximations to calculate the averaged fidelity of quantum teleportation in the
ultraweak coupling limit.

\begin{figure}
\includegraphics[width=5.5cm]{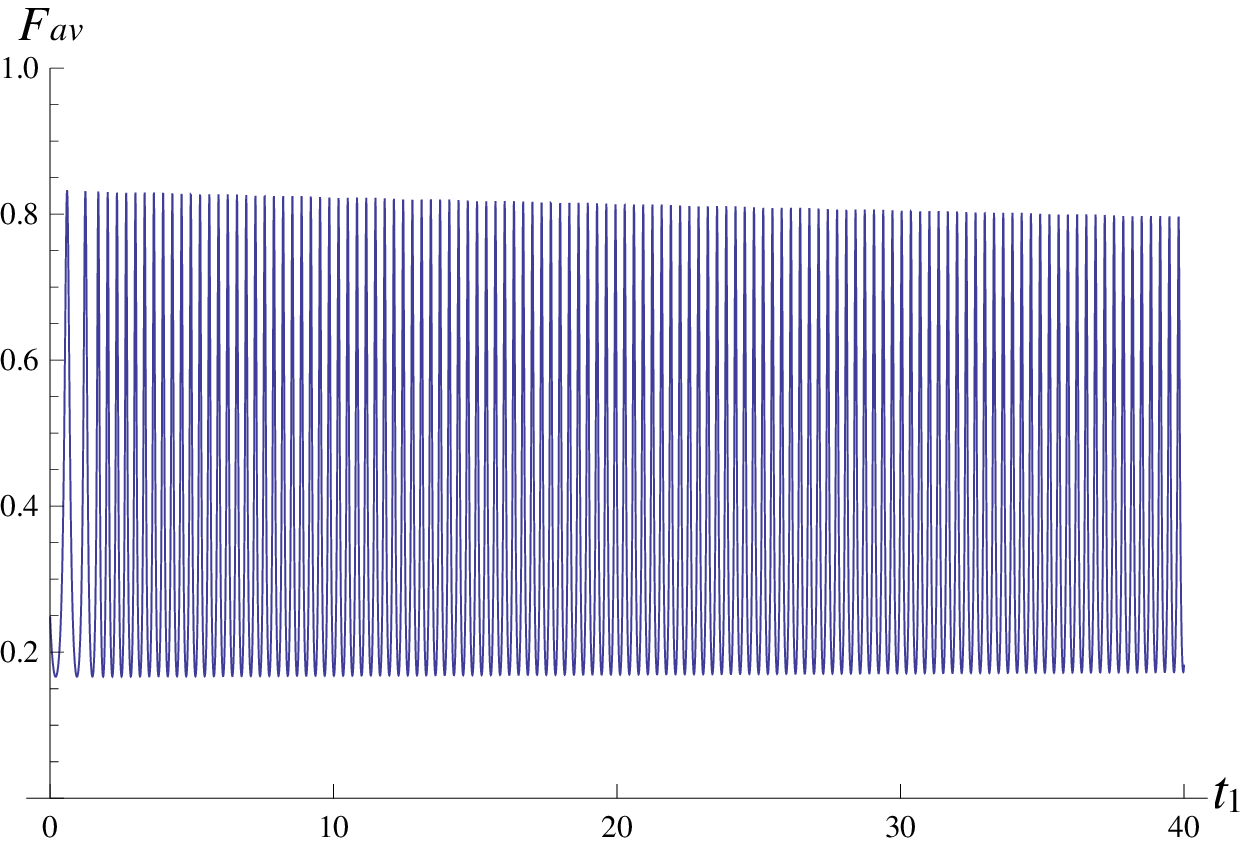}
\includegraphics[width=5.5cm]{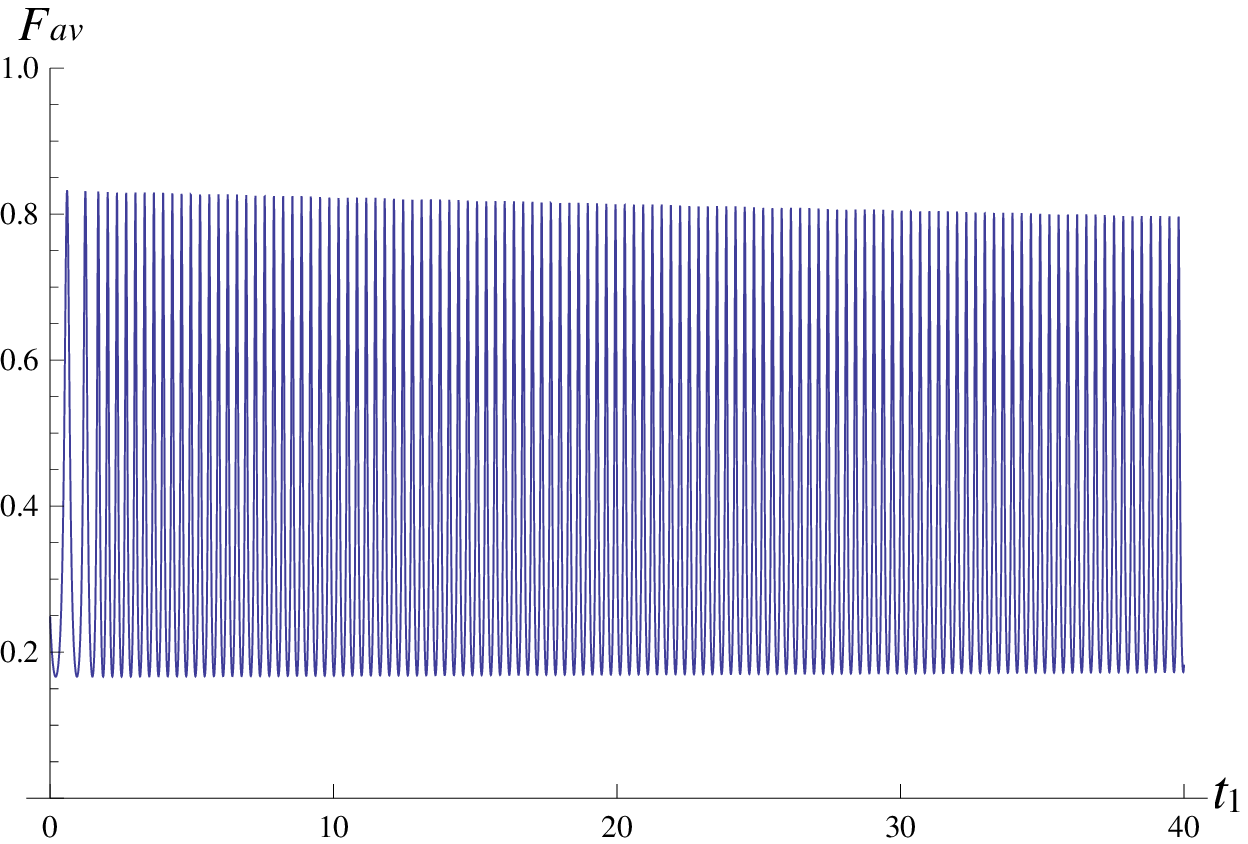}
\includegraphics[width=5.5cm]{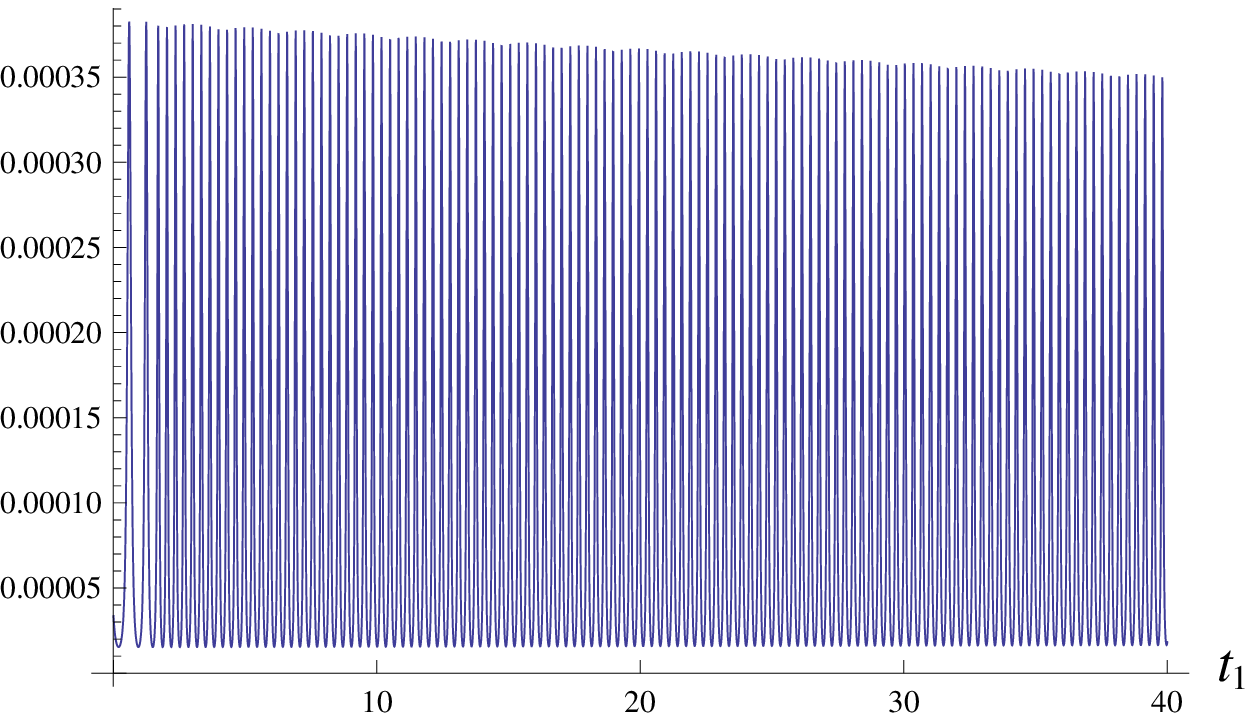}
\caption{Averaged physical fidelity $F_{av}$ against the moment $t_1$ of the joint measurement on $A$ and $C$ {\it in the ultraweak coupling
limit} in the more realistic cases where the detector $B$ stops accelerating at $\tau_2=10$, and the local operation on detector $B$ is performed
right after $\tau^{adv}_1$. Before this moment the wave functional was collapsed either on Minkowski time-slice then evolve to
$\tau_1^{adv}$ (left) or almost on the future lightcone of the joint measurement event by Alice (middle), namely,
the hypersurface intersecting the worldlines of Alice and Rob at $t_1$ and $\tau_1^{adv}$.
Other parameters are the same as those in Fig. \ref{wcCorr}.
The difference between the results in the left and the middle plots
is shown in the right plot. One can see that it is within $O(\gamma)$ (here $\gamma=0.0002$).
Compare the left plots with Fig. \ref{wcF} (left) one can see the huge
difference from the averaged pseudo-fidelity there. The difference is due to the natural oscillations of the
detector $B$ from $t_1$ to the much later moment $\tau_1^{adv}$.}
\label{RealWeak}
\end{figure}

\subsection{Averaged physical fidelity of quantum teleportation in ultraweak coupling limit}

\begin{figure}
\includegraphics[width=5.5cm]{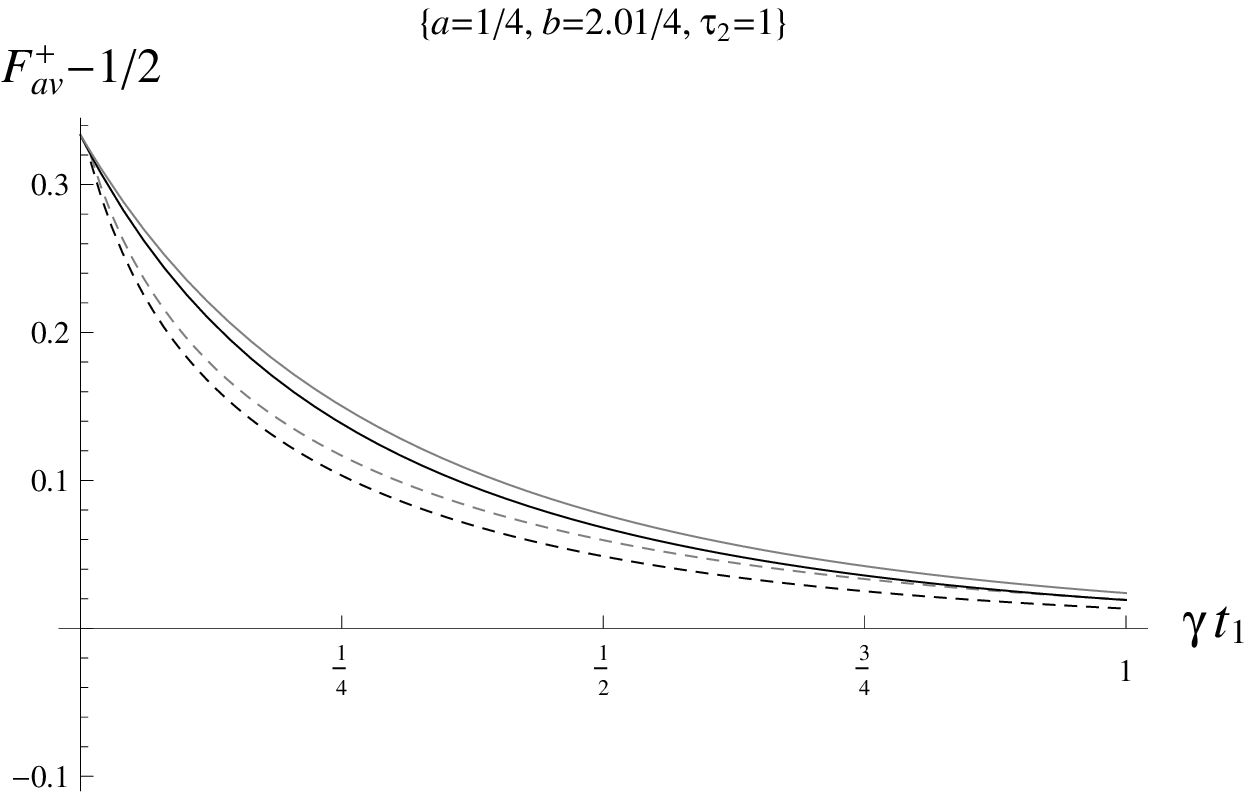}
\includegraphics[width=5.5cm]{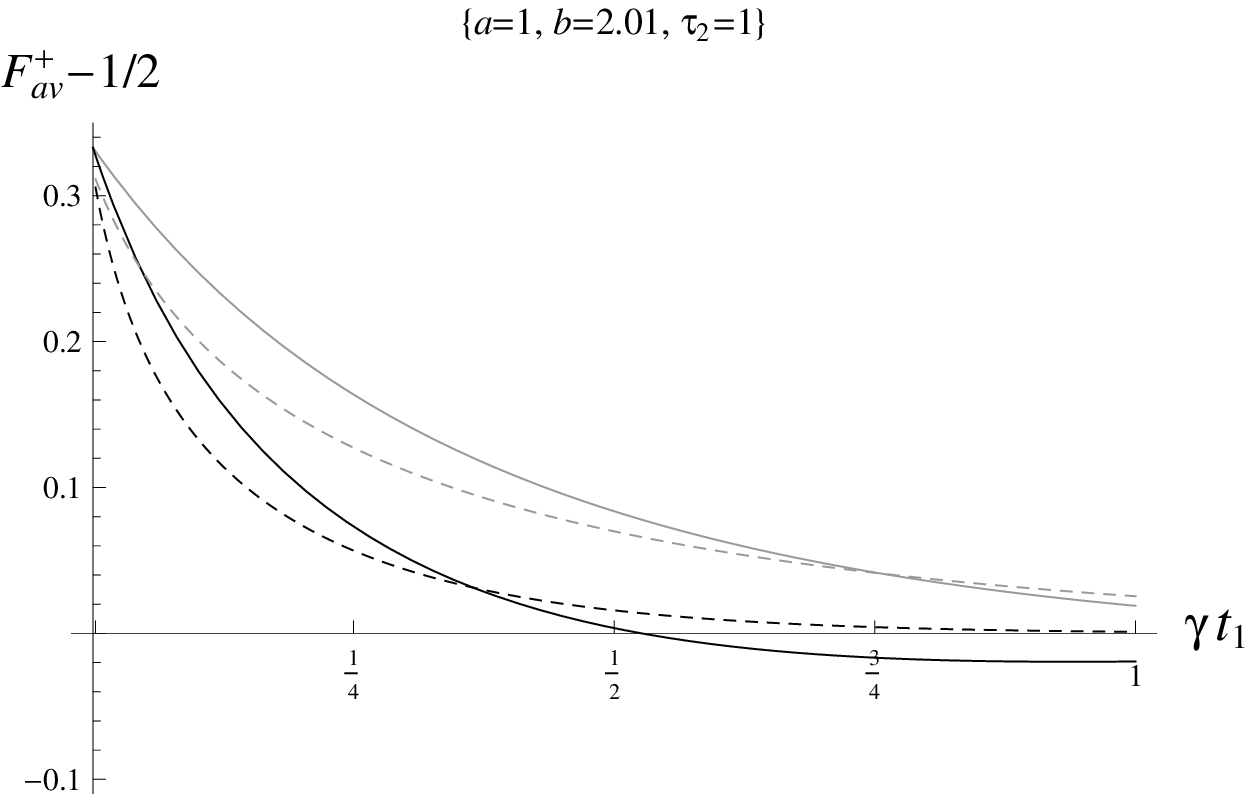}
\includegraphics[width=5.5cm]{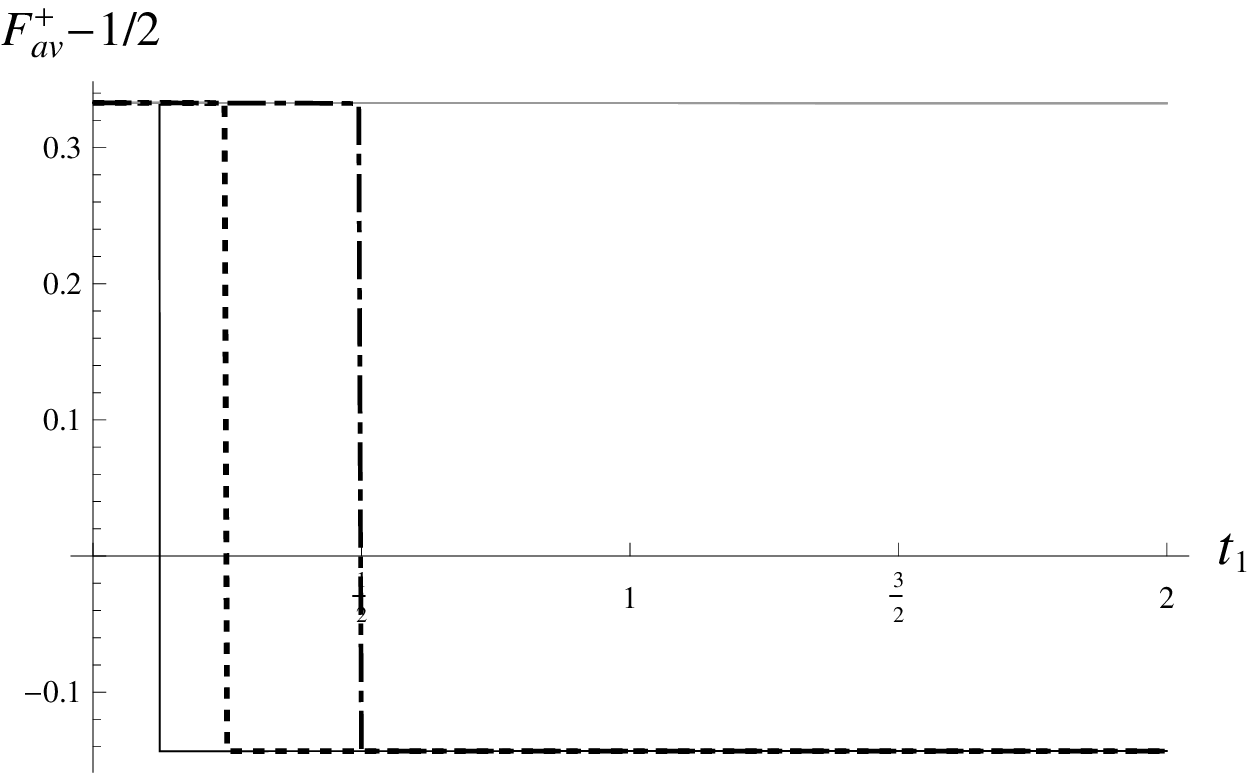}
\includegraphics[width=5.5cm]{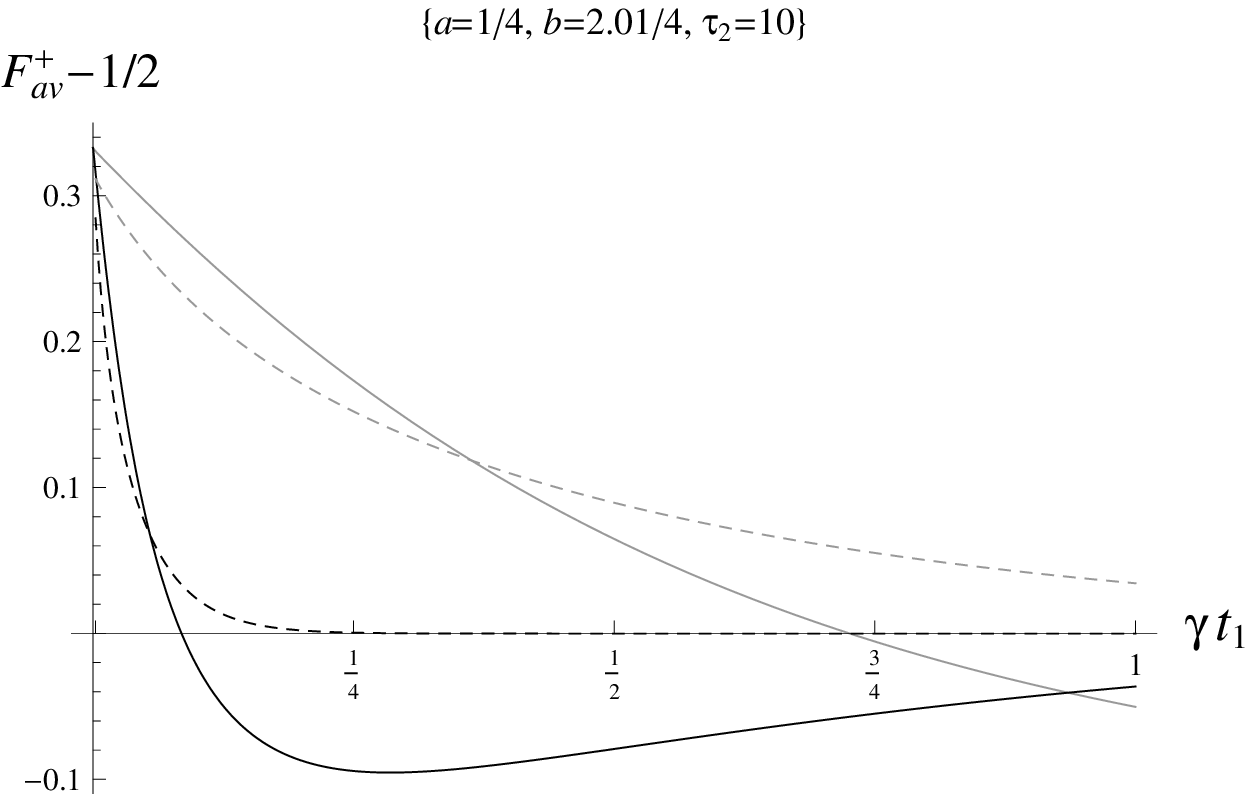}
\includegraphics[width=5.5cm]{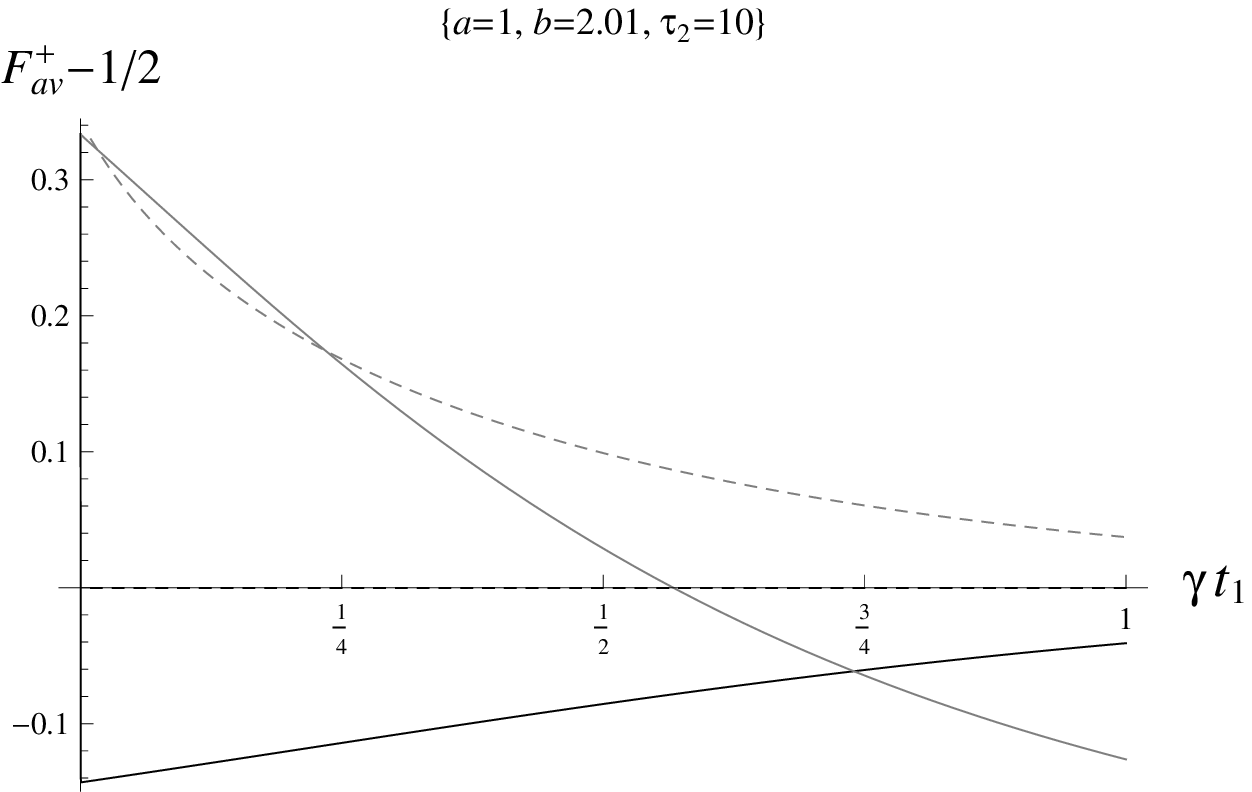}
\includegraphics[width=5.5cm]{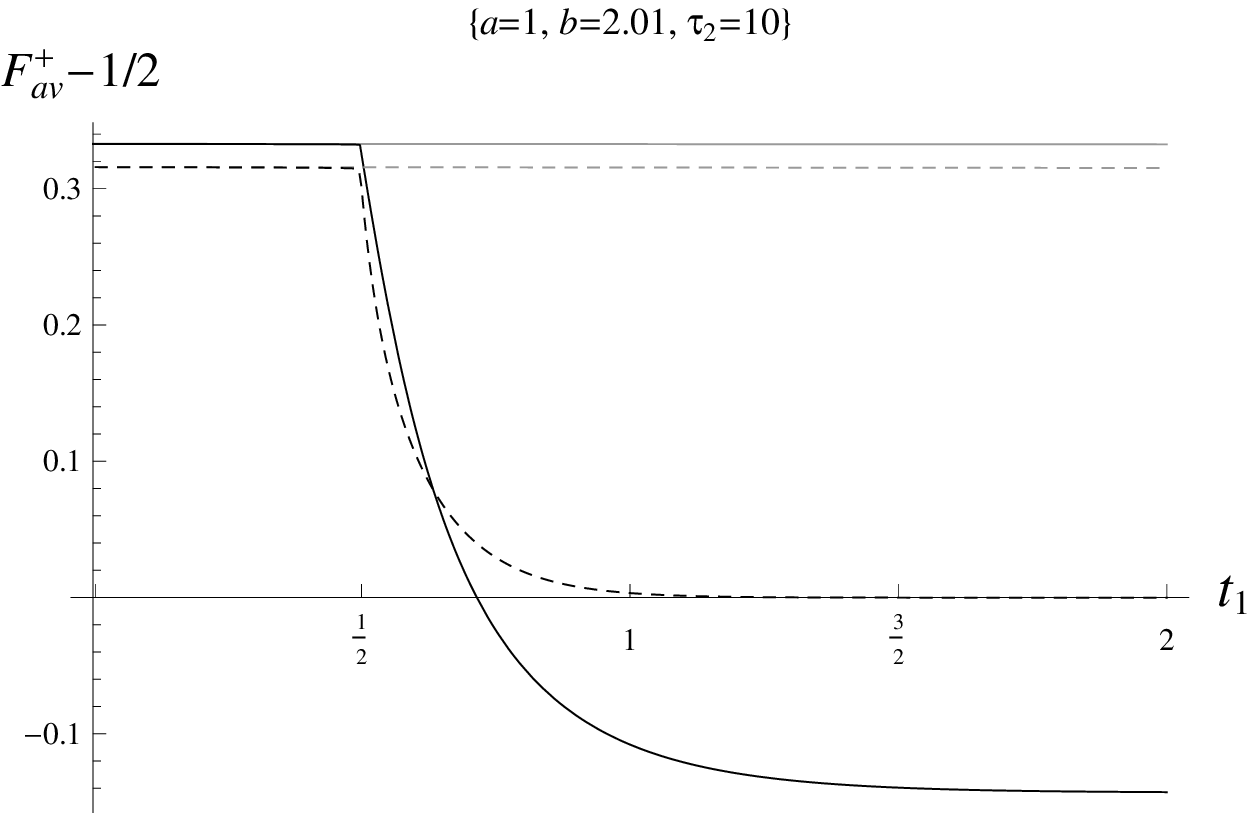}
\caption{$F_{av}^+-1/2$ (solid curves) and $E_{\cal N}/10$ (dashed curves) in the Minkowski frame in the more realistic case
as a function of the moment of the joint measurement $t_1$.
The black curves are those physical ones with the wave functionals almost collapsed on the lightcone,
while the gray curves are those pseudo-fidelities with the wave functionals collapsed on the $t_1$-slice in the Minkowski frame.
Here the parameters are the same as those in Fig. \ref{wcCorr} except $a$, $b$ and $\tau_2$.
One can see that the moment that the peak values of the averaged physical fidelity $F_{av}^+$ becomes less than $1/2$ is always earlier
than the pseudo-$F_{av}^+$ has, and both happens earlier than the disentanglement times evaluated on the corresponding 
hypersurfaces of wave functional collapse.
The lower-right plot is a close up of the lower-middle plot at very early times. If we increase
the value of $a\tau_2$ further we get the upper-right plot, where
the solid black curve indicates the case with $a=4$, $b=8.04$, $\tau_2=10$,
the dotted curve indicates the case with $a=2$, $b=4.02$, $\tau_2=10$,
and the dot-dashed curve indicates the case with $a=1$, $b=2.01$, $\tau_2=20$.
All these three curves become negative at about $t_1=1/b$.}
\label{FavPlus}
\end{figure}

Replacing $\tau_1$ in $(\ref{tildeVwc})$ by $\tau^{adv}_1$ in $(\ref{tau1adv})$,
while inserting $(\ref{QB2NUAD})$ and $(\ref{PB2NUAD})$ into the v-parts of the self correlators in ${\cal Q}_{BB}^{[1]}$
and ${\cal P}_{BB}^{[1]}$, respectively, we obtain the results in Figs. \ref{RealWeak}(middle) and \ref{FavPlus}.

In Fig. \ref{RealWeak}(right) one can see that the differences between the results with wave functional collapsed on Minkowski time-slice
then evolving the system to $\tau_1^{adv}$ (left), and those collapsed almost on the future lightcone of the measurement event
(middle), are $O(\gamma)$, which is within the error of the two-point correlators in the ultraweak coupling limit so they should be
considered negligible.

In all the plots of Fig. \ref{RealWeak} the number of peaks of the physical $F_{av}$ in the same duration of $t_1$ in this more
realistic case is much more than the one for the averaged pseudo-fidelity. This is because it takes a long time from $\tau_1$ to
the moment $\tau_1^{adv}$ when the classical signal from Alice reaches Rob, during which detector $B$ has oscillated many times.

In Fig. \ref{FavPlus} we see that the moment $t_1=t_{1/2}$ when the best averaged physical fidelity of quantum
teleportation $F_{av}^+$ drops to $1/2$ is again earlier than any $F_{av}^+$ of pseudo-fidelity has.
So Alice must perform the joint measurement on detectors $A$ and $C$ much earlier
than what was estimated from the pseudo-fidelities to achieve successful quantum teleportation.
The larger $a\tau_2$, the later $\tau^{adv}_1$ Rob has by $(\ref{tau1adv})$,
and so the lower value of the physical $F_{av}^+$ at that time due to the longer time
of coupling with the environment. When $a\tau_2$ is large enough, $\tau_1^{adv}$ is so large that $t_{1/2}\approx b^{-1}$, which is the moment
that Alice enters the event horizon of Rob for $\tau_2\to\infty$ 
(see Fig. \ref{FavPlus} (upper-right), (lower-middle), and (lower-right)).

From the same argument in Appendix \ref{FavEntUwc} with the proper time of detector $B$ substituted
by $\tau_1^{adv}$ (actually $\tau_1^{adv}\pm \epsilon$ as $\epsilon\to 0+$), one can see that quantum entanglement of $AB$-pair
evaluated almost on the future lightcone of the measurement event by Alice is still a necessary condition of
the best averaged physical fidelity of quantum teleportation beating the classical one in the ultraweak coupling limit
of our model.

\section{Summary}

Quantum entanglement of two localized but spatially separated objects is a kind of spacelike correlations while the physical fidelity of quantum teleportation, is a kind of timelike correlations. In general they are incommensurate. 
We define the pseudo-fidelity of quantum teleportation on the same time-slice that the joint measurement occurs
in some reference frame to compare with the quantum entanglement of the $AB$-pair on that time-slice.
In the more realistic cases assuming that Rob stops accelerating after some $\tau_2 > 0$,
the classical signal from Alice can always reach Rob, and
the reduced state of detector $B$ collapsed on different time-slices by the same joint measurement by Alice
in different frames will become consistent when Rob is entering the future lightcone of the measurement event.
Thus we are allowed to perform the projection of the wave functional almost on that future lightcone, and right
after that, let Rob perform the local operation. The physical fidelity of quantum teleportation obtained
in this way can be compared with the quantum entanglement of the $AB$-pair evaluated right before the wave functional
collapsed almost on the future lightcone of the joint measurement event by Alice.

In the ultraweak coupling limit of our model, when detectors $A$ and $B$ are separable on a time-slice,
the averaged pseudo-fidelity $F_{av}$ of quantum teleportation obtained on that time-slice must be less
than or equal to $1/2$, which is the best possible fidelity of classical teleportation.
Thus quantum entanglement between detectors $A$ and $B$ is a necessary condition that the averaged
pseudo-fidelity of quantum teleportation has advantage over the best classical one.
We have a similar observation in our result beyond the ultraweak coupling limit.

Similarly, if we assume that the classical signal from Alice travels with lightspeed and Rob performs the local
operation right after he received the signal, then in the ultraweak coupling limit the entanglement ``on the lightcone",
which is evaluated right before Rob enters the future lightcone of the measurement event by Alice, will also be a necessary
condition for a physical fidelity of quantum teleportation beating the classical ones.

We have seen that the logarithmic negativity $E_{\cal N}$ evolves quite smoothly while the averaged
pseudo-fidelity evaluated in whatever reference frame or the averaged physical fidelity of quantum teleportation oscillates
in $t_1$, which is the moment when Alice performs the joint measurement on detectors $A$ and $C$. Even at very early times
$F_{av}$ drops below $1/2$ frequently. It is clear that the oscillation of the averaged fidelities are mainly
due to the distortion of the quantum state of the $AB$-pair from their initial state
(caused by the alternating squeeze-antisqueeze natural oscillations of their quantum state)
rather than the time evolution of quantum entanglement between them.

In all cases  considered in this paper different values of $a$ have different ways of time-dilation,
which causes different shifts of the peaks of $F_{av}$ at early times in $t_1$, though the $a$-dependence of
the peak values of $F_{av}$ is not significant at this stage in all cases in the ultraweak coupling limit.
In a longer time scale, while the peak values of the averaged pseudo-fidelity in the Minkowski frame are insensitive to the
proper acceleration $a$, the ones in the quasi-Rindler frame as well as
the averaged physical fidelity in the more realistic cases do depend on $a$ significantly:
the larger $a$ is, the quicker the best fidelities $F_{av}^+ (t_1)$ drops below $1/2$.

Finally,
the best averaged physical fidelity becomes less than $1/2$ always earlier (in $t_1$) than any best averaged pseudo-fidelity
we considered does, while each of these moments is earlier than the disentanglement time evaluated on the corresponding
hypersurfaces which the wave functional collapses on.
In our more realistic cases, the later Rob turns into inertial motion (i.e. the larger $a \tau_2$),
the earlier the moment in $t_1$ that 
the best averaged physical fidelity drops below $1/2$, and the earlier the corresponding disentanglement time in $t_1$.
For $a \tau_2$ large enough, detectors $A$ and $B$ become separable ``on the lightcone" right after $t_1=1/b$
in the ultraweak coupling limit, meaning that quantum teleportation loses advantage right after Alice goes beyond the
event horizon of Rob in the limiting case $\tau_2 \to \infty$ in our setup.
\\

\begin{acknowledgments}
Part of this work was done while BLH  visited the National Center for Theoretical Sciences (South) and the Department of Physics of National Cheng-Kung University, Tainan, Taiwan, the Center for Quantum Information and Security at Macquarie University, Sydney,  the Center for Quantum Information and Technology at the University of Queensland, Brisbane, Australia in January-March, 2011 and the National Changhua University of Education, Taiwan in January 2012. He wishes to thank the hosts of these institutions for their warm hospitality.   This work is supported by the National Science Council of Taiwan under grant NSC 100-2112-M-006-007, NSC 99-2112-M-018-001-MY3, and in part by the National Center
for Theoretical Sciences, Taiwan, and by USA NSF PHY-0801368 to the University of Maryland.
\end{acknowledgments}

\begin{appendix}

\section{Averaged fidelity and entanglement in ultraweak coupling limit}
\label{FavEntUwc}

In our model in the ultraweak coupling limit, it is straightforward to show to the leading order that 
quantum entanglement between detectors $A$ and $B$ is a necessary condition for the corresponding 
averaged fidelity of quantum teleportation to be better than the best classical ones. 

Suppose in the post-measurement state
$\tilde{\cal Q}_{AA}=\tilde{\cal Q}_{CC}= \hbar C_2/2\Omega$, $\tilde{\cal Q}_{AC}=\hbar S_2/ 2\Omega$,
$\tilde{\cal P}_{AA}=\tilde{\cal P}_{CC}=\hbar\Omega C_2/2$, $\tilde{\cal P}_{AC}=\hbar\Omega S_2/2$, 
and $\tilde{\cal R}_{mn}=0$ (with $C_n \equiv \cosh 2r_n$ and $S_n\equiv \sinh 2r_n$, see Section \ref{infiweak}).
Also from Eqs.$(28)$, $(29)$, $(32)$, $(33)$, and $(B2)$-$(B8)$ in Ref. \cite{LCH08} with $\alpha^2 = (\hbar/\Omega)e^{-2r_1}$
and $\beta^2 = \hbar\Omega e^{-2r_1}$ there (not the complex numbers $\alpha$ and $\beta$ in this paper),
writing $\upsilon \equiv 2\hbar\gamma\Lambda_1/\pi$, with
$1\gg \upsilon \gg \gamma \gg \upsilon^2$, one has
\begin{eqnarray}
 {\cal Q}_{AA} &\approx&{\hbar\over 2\Omega}\left[ C_1 e^{-2\gamma t_1} + 1-e^{-2\gamma t_1}\right]+O(\gamma),\label{QAAwc}\\
 {\cal Q}_{BB} &\approx& {\hbar\over 2\Omega}\left[ C_1 e^{-2\gamma \tau_1} + 1-e^{-2\gamma \tau_1}\right] +
                         \delta\left<\right. \hat{Q}_B^2(\tau_1)\left.\right>_{\rm v}+O(\gamma),\\
 {\cal Q}_{AB} &=& {\hbar\over 2\Omega} S_1 e^{-\gamma(t_1+\tau_1)}\cos\Omega(t_1+\tau_1)+O(\gamma),\\
 {\cal P}_{AA} &\approx& {\hbar\Omega\over 2}\left[ C_1 e^{-2\gamma t_1} + 1-e^{-2\gamma t_1}\right] + \upsilon +O(\gamma),\\
 {\cal P}_{BB} &\approx& {\hbar\Omega\over 2}\left[ C_1 e^{-2\gamma \tau_1} + 1-e^{-2\gamma \tau_1}\right] + \upsilon+
                         \delta\left<\right. \hat{P}_B^2(\tau_1)\left.\right>_{\rm v}+O(\gamma),\\
 {\cal P}_{AB} &\approx& -\Omega^2 {\cal Q}_{AB}+O(\gamma), \\
 {\cal R}_{AB} &\approx& {\cal R}_{BA}
   \approx {\cal R} \equiv -{\hbar\over 2}S_1 e^{-\gamma(t_1+\tau_1)}\sin\Omega(t_1+\tau_1)+O(\gamma), \\
 {\cal R}_{AA} &\approx& {\cal R}_{BB} \approx 0+O(\gamma), \label{RAAwc}
\end{eqnarray}
for the initial state $(\ref{rhoABI})$ in the ultraweak coupling limit. Here
$\delta\left<\right. \hat{P}_B^2(\tau_1)\left.\right>_{\rm v}\approx \Omega^2 \delta\left<\right. \hat{Q}_B^2(\tau_1)
\left.\right>_{\rm v} \approx (\hbar/2\Omega)(\coth(\pi\Omega/a)-1)(1-e^{-2\gamma \tau_1})$ if Rob is uniformly accelerated
($\tau_2\to \infty$) with proper acceleration $a$, and $\delta\left<\right. \hat{Q}_B^2(\tau_1)\left.\right>_{\rm v} \equiv
\left<\right. \hat{Q}_B^2(\tau_1)\left.\right>_{\rm v}-\left<\right. \hat{Q}_B^2(\tau_1)|_{a_\mu a^\mu \to 0}\left.\right>_{\rm v}$
and $\delta\left<\right. \hat{P}_B^2(\tau_1)\left.\right>_{\rm v}\equiv \left<\right. \hat{P}_B^2(\tau_1)\left.\right>_{\rm v}-
\left<\right. \hat{P}_B^2(\tau_1)|_{a_\mu a^\mu \to 0}\left.\right>_{\rm v}$ $(\approx \Omega^2 \delta\left<\right. \hat{Q}_B^2(\tau_1)
\left.\right>_{\rm v}$, too) are given in $(\ref{QB2NUAD})$ and $(\ref{PB2NUAD})$ in the more realistic case, so
\begin{equation}
  {\cal P}_{AA}\approx\Omega^2 {\cal Q}_{AA}+\upsilon + O(\gamma), \hspace{.5cm}
  {\cal P}_{BB}\approx\Omega^2 {\cal Q}_{BB}+\upsilon + O(\gamma).
\end{equation}
Then on the time-slices passing through the worldlines of detectors $A$ and $B$ at $t_1$ and $\tau_1$, respectively,
\begin{eqnarray}
  \Sigma &\approx& {\cal Z}^2 -{\hbar^2\over 4}\Omega^2\left({\cal Q}_{AA}+{\cal Q}_{BB}\right)^2 +
    \upsilon ({\cal Q}_{AA}+{\cal Q}_{BB} )  \left({\cal Z}-{\hbar^2\over 2}\right)+O(\gamma)\\
  \hbar^2\pi N_B &\approx& \hbar \sqrt{ \left(\Omega{\cal Q}_{AA}+{\hbar\over 2}C_2\right)^2 + \upsilon
    \left({\cal Q}_{AA}+{\hbar\over 2\Omega}C_2\right)}+O(\gamma), \\
  \det \tilde{V} &\approx& {\cal F}^2+ \upsilon\left[ {\cal Q}_{AA}+{\cal Q}_{BB}+ {\hbar\over \Omega}(1+C_2)
    \right]{\cal F} + O(\gamma),
\end{eqnarray}
where $\Sigma$ defined in \cite{LCH08} indicates the degree of entanglement between detectors $A$ and $B$,
${\cal Z}$ and ${\cal F}$ are given by
\begin{eqnarray}
  {\cal Z} &\equiv& {\hbar^2\over 4} + \Omega^2{\cal Q}_{AA}{\cal Q}_{BB} -\Omega^2 {\cal Q}_{AB}^2-{\cal R}^2,\\
  {\cal F} &\equiv& {\cal Z} + \hbar\Omega{\cal Q}_{AA} - \hbar S_2 \Omega{\cal Q}_{AB}+
    {\hbar\over 2} C_2\left[\hbar + \Omega ({\cal Q}_{AA}+{\cal Q}_{BB})\right].
\end{eqnarray}
Detectors $A$ and $B$ are separable if and only if $\Sigma \ge 0$, or
\begin{equation}
  {\cal Z} \ge {\hbar\over 2}\Omega\left({\cal Q}_{AA}+{\cal Q}_{BB}\right) -{\upsilon\over 2}
  \left({\cal Q}_{AA}+{\cal Q}_{BB}-{\hbar\over \Omega}\right) + O(\gamma).
\end{equation}
This implies
\begin{eqnarray}
  & &\det\tilde{V}-4\left(\hbar^2\pi N_B\right)^2 \ge {\cal J}^2 - \hbar^2\left(2\Omega{\cal Q}_{AA}+\hbar C_2\right)^2 +
  \nonumber\\ & & \,\,\,\,\,\upsilon{\cal J}\left[ {3\over 16}\left({\cal Q}_{AA}+{\cal Q}_{BB}\right)+
  {\hbar\over \Omega}\left({5+4C_2\over 16}\right)\right]
  -2\upsilon\hbar^2\left(2{\cal Q}_{AA} + {\hbar\over\Omega} C_2\right)
  + O(\gamma), \label{Favineq}
\end{eqnarray}
where
\begin{equation}
  {\cal J} \equiv {\hbar\over 2}\Omega (1+C_2)\left({\cal Q}_{AA}+{\cal Q}_{BB}\right) +\hbar\Omega{\cal Q}_{AA}
    -\hbar S_2 \Omega{\cal Q}_{AB}+ {\hbar^2\over 2} C_2.
\end{equation}
Now one can see 
\begin{eqnarray}
  & & {\cal J} - 2\hbar \left(\Omega{\cal Q}_{AA}+{\hbar\over 2}C_2\right)\nonumber\\
  &=& \hbar\Omega \left[ \left({C_2-1\over 2}\right)\left( {\cal Q}_{AA}-{\hbar\over 2\Omega}\right) +
    \left({C_2+1\over 2}\right)\left( {\cal Q}_{BB}-{\hbar\over 2\Omega}\right)-S_2 {\cal Q}_{AB}\right] \nonumber\\
  &\ge& \hbar^2 \left(\sinh r_1 \sinh r_2 e^{-\gamma t_1}+\cosh r_1 \cosh r_2 e^{-\gamma\tau_1} \right)^2 +
    \hbar\Omega \,\,\delta\left<\right.\hat{Q}_B^2(\tau_1)\left.\right>_{\rm v} \cosh^2 r_2>0,
  \label{Jineq}
\end{eqnarray}
after the approximated expressions for the correlators were inserted. Notice that $\delta\left<\right.\hat{Q}_B^2(\tau_1)
\left.\right>_{\rm v}$ is always positive here. This implies that the $O(\upsilon^0)$
terms of $\det\tilde{V}-4(\hbar^2\pi N_B)$ in $(\ref{Favineq})$ is positive whenever $\Sigma\ge 0$, and
the $O(\upsilon)$ terms in $(\ref{Favineq})$ in this case must be greater than
\begin{equation}
  8\upsilon\hbar \left(\Omega{\cal Q}_{AA}+{\hbar\over 2}C_2\right)\left[ {3\over 16}
  \left({\cal Q}_{AA}+{\cal Q}_{BB}\right)+ {\hbar\over \Omega}\left({1+4C_2\over 16}\right)\right] >0. \nonumber
\end{equation}
Therefore, for the entangled pair of the UD detectors initially in the state $(\ref{rhoABI})$
in the ultraweak coupling limit, if $\Sigma \ge 0$, then $\det \tilde{V} > 4(h^2\pi N_B)^2$ up to
$O(\hbar\gamma\Lambda_1)$ and so $F_{av} = h^2\pi N_B / \sqrt{\det\tilde{V}}$ must be less than $1/2$ to this order.
In other words, once quantum entanglement between $A$ and $B$ disappears, the corresponding averaged pseudo-fidelity of quantum
teleportation must have been less than the best classical fidelity $1/2$ in the ultraweak coupling limit of our model.

\section{Entanglement, quantum nonlocality, and causality}

Suppose $\tau_2\to\infty$ so Rob has an event horizon at the hypersurface $t=x$. Suppose 
at the moment $t_1$ before Alice goes beyond the event horizon of Rob (Fig. \ref{AR}) Alice performs a joint measurement
on $A$ and $C$. Then the wave functional of the combined system (including the quantum fields that these UD detectors are coupled with)
will collapse at that moment either on the Minkowski time-slice (gray dashed horizontal line in Fig. \ref{AR}), or on the quasi-Rindler
time-slice (the gray solid curve in the same plot), 
or whatever time-slice may be passing through the same measurement event in some observer's frame.
All  post-measurement states of the combined system in different frames will evolve to
the same state when compared on the same time-slice after the measurement \cite{Lin11a}.

The Gaussian reduced state of the detector $B$ consists of the two-point correlators of $Q_B$ and $P_B$. It has a sudden change from
the uncollapsed to the collapsed one at $\tau^{}_B=\tau_1$ in Fig. \ref{AR}, as observed in the conventional Minkowski frame,
or at $\tau'_1$, as observed in the quasi-Rindler frame. Such a sudden change occurs at different spacetime points for different observers.
In other words, when observed at some moment $\tau^{}_B < \tau^{adv}_1$ before Rob enters the future lightcone of the joint measurement event,
those correlators of $B$ may either be in the uncollapsed form or the collapsed form, with the two-point correlators look like
$(\ref{QA2examp})$ or $(\ref{QB2clpsed})$, depending on the observer \cite{Lin11b}.
In both cases all the correlators of $B$ are independent of the data on the
time-slice that the wave functional collapsed on except those localized right at the position of the detector $A$ and $C$.

If Rob never performs any further measurement on $B$ before entering the future lightcone of the joint measurement event
done by Alice, certainly he will have no idea that $B$ is in the uncollapsed or collapsed state.
But right before the moment $\tau^{}_B =\tau^{adv}_1$ when Rob is entering the future lightcone of the joint measurement event,
these different reduced states of $B$ in different frames must have become the same collapsed one, with the same
combination of the mode functions depending only on the data on the initial time-slice.
So quantum teleportation after Rob receives the classical information from Alice will give a definite result consistent in all frames.

Suppose Rob performs a measurement on $B$ before entering the future lightcone of the joint measurement event to see which
state $B$ is in. In some observers' frames $B$'s reduced state at the moment of Rob's measurement is in the collapsed state, then the
outcome will have some dependence on the outcome of $A$ since $A$ and $B$ were entangled initially. In other observers' frames,
$B$ is still in the uncollapsed state before the measurement, then this measurement will result in a wave functional
collapse of the combined system, later in these frames the outcome of Alice's joint measurement will have the same correlation
with the outcome of $B$. Both kinds of histories interpreted by different observers will be consistent {\it a posteriori},
but both could not help Rob to conclude $B$ was in the uncollapsed or collapsed state right before the measurement
because the quantum state is only sampled by Rob in one single measurement.
If Rob and Alice share an ensemble of many copies of the entangled pairs, in some observers' frames Rob performs the measurement
first, then using the outcomes Rob can recognize the reduced state of $B$ as uncollapsed by quantum state tomography. In other
observers' frames, Rob performs the measurement after Alice. For these observers the exact collapsed state of one copy of the entangled
pair may be different from another after the measurements by Alice and not predictable (either by Rob or by Alice), while the distribution
of the collapsed state is determined by the uncollapsed state right before Alice's measurement. So the reduced state of detector
$B$ recognized by Rob from his measurement on this ensemble of the collapsed states will have no difference from the uncollapsed one.
Rob still cannot determine whether the reduced state of the {\it single} detector $B$ before his measurement is collapsed or not.

If the joint measurement on $A$ and $C$ is performed after Alice went beyond the event horizon $t=x$ of Rob,
the mutual influences ($\phi_B^A$ and $f_B^A$ in $(\ref{Updef})$) will never reach Rob,
though it appears that some information of measurement
could enter the collapsed reduced state of $B$ \cite{Lin11b}. Similar to the previous case with Rob still outside
the future lightcone of the joint measurement, the functional form of the correlators in the reduced state of $B$
will be suddenly changed at the moment of the projective measurement,
and different observers will recognize different spacetime points where this change occurs.
But again Rob will never know whether the state of $B$ is in the uncollapsed or collapsed state.
These reduced states in different frames will not be identical until $\tau^{}_B$ goes to infinity or
some moment Rob performs a measurement on $B$.

\end{appendix}

\end{document}